\documentclass[prc,floatfix,groupedaddress,nofootinbib,showpacs,preprintnumbers,
amsmath,amssymb,amsfonts,superscriptaddress,widetable] {revtex4}
\usepackage{graphicx}
\usepackage{dcolumn}
\usepackage{bm}
\usepackage{mathrsfs}
\usepackage[usenames]{color}
\begin{document}

\title{Pygmy Resonances and Neutron Skins} 
\author{J. Piekarewicz}
\affiliation{Department of Physics, Florida State 
             University, Tallahassee, FL 32306}
\date{\today} 

\begin{abstract}
 Motivated by a recent experiment, the distribution of electric dipole
 strength in the neutron-rich ${}^{68}$Ni isotope was computed using a
 relativistic random phase approximation with a set of effective
 interactions that---although well calibrated---predict significantly
 different values for the neutron-skin thickness in ${}^{208}$Pb.  The
 emergence of low-energy ``Pygmy'' strength that exhausts about 5-8\%
 of the energy weighted sum rule (EWSR) is clearly identified. In
 addition to the EWSR, special emphasis is placed on the dipole
 polarizability. In particular, our results suggest a strong correlation
 between the dipole polarizability of ${}^{68}$Ni and the neutron-skin
 thickness of ${}^{208}$Pb. Yet we find a correlation just as strong
 and an even larger sensitivity between the neutron-skin thickness of
 ${}^{208}$Pb and the {\sl fraction} of the dipole polarizability
 exhausted by the Pygmy resonance. These findings suggest that the
 dipole polarizability may be used as a proxy for the neutron skin.
\end{abstract}
\maketitle 

\section{Introduction}
\label{Sec:Intro}
The determination of the neutron radius of a heavy nucleus is a
problem of fundamental importance that has been raised to a place of
prominence due to its far reaching implication in areas as diverse as
atomic parity violation~\cite{Pollock:1992mv,Sil:2005tg}, nuclear
structure~\cite{Brown:2000,Furnstahl:2001un,Danielewicz:2003dd,
Centelles:2008vu,Centelles:2010qh},
heavy-ion collisions~\cite{Tsang:2004zz,Chen:2004si,Steiner:2005rd,
Shetty:2007zg,Tsang:2008fd}, and neutron-star
structure~\cite{Horowitz:2000xj,Horowitz:2001ya,Horowitz:2002mb,
Carriere:2002bx,Steiner:2004fi,Li:2005sr}. Earlier attempts at mapping
the neutron distribution of heavy nuclei were met with skepticism as
they relied on strongly-interacting processes that are handicapped by
large and controversial uncertainties in the reaction
mechanism~\cite{Ray:1985yg,Ray:1992fj}. Instead, the enormously
successful parity-violating program at the Jefferson Laboratory
provides an attractive electroweak alternative.  The Lead Radius
experiment (PREx) aims to determine the neutron radius of $^{208}$Pb
accurately and model independently using parity-violating electron
scattering~\cite{Horowitz:1999fk,Michaels:2005}.  Parity violation at
low momentum transfers is particularly sensitive to the neutron
distribution because the neutral weak-vector boson $Z^0$ couples
preferentially to the neutrons. Moreover, the parity-violating
asymmetry, although very small, may be interpreted with as much
confidence as conventional electromagnetic scattering experiments that
have been used for decades to map the proton distribution with
exquisite accuracy. The Lead Radius experiment was successfully
commissioned in March of 2010. High quality data were collected at the
designed luminosity and with sufficient statistics to provide (likely
by the Spring of 2011) a significant first experimental constraint on
the neutron radius of $^{208}$Pb~\cite{Michaels:2005}.

A promising complementary approach to the parity-violating program
relies on the electromagnetic excitation of the electric dipole
mode~\cite{Harakeh:2001}.  For stable (medium to heavy) nuclei with a
moderate neutron excess the dipole response is concentrated on a
single fragment---{\sl the giant dipole resonance (GDR)}---that
exhausts almost 100\% of the Thomas-Reiche-Kunz (TRK) sum rule. (The
TRK sum rule is proportional to the energy-weighted sum rule (EWSR) so
we will use both terms interchangeably.)  For this mode of
excitation---perceived as an oscillation of neutrons against
protons---the symmetry energy acts as its restoring force.  Models
with a soft symmetry energy, namely, ones that change slowly with
density, predict large values for the symmetry energy at the densities
of relevance to the excitation of this mode. As a consequence, the
stronger restoring force of the softer models generates a dipole
response that is both hardened ({\sl i.e.,} pushed to higher
excitation energies) and quenched relative to its stiffer
counterparts. Given that the neutron radius of a heavy nucleus is also
sensitive to the density dependence of the symmetry energy, the peak
position of the GDR may be used as a (mild) constrain on the neutron
radius.

A more stringent constrain on the neutron radius is expected to emerge
as the nucleus develops a neutron-rich skin. Concomitant with the
development of a neutron skin is the appearance of low energy dipole
strength---the so-called {\sl pygmy dipole resonance
(PDR)}~\cite{Suzuki:1990,VanIsacker:1992,Hamamoto:1996,
Hamamoto:1998,Vretenar:2000yy,Vretenar:2001hs,Paar:2004gr}.  Thus, it
has been suggested that the PDR---perceived as an excitation of the
neutron-rich skin against the symmetric core---may be used as a
constraint on the neutron skin of heavy
nuclei~\cite{Piekarewicz:2006ip}.  In particular, the {\sl fraction}
of the EWSR exhausted by the pygmy resonance has been shown to be
sensitive to the neutron-skin thickness of heavy
nuclei~\cite{Tsoneva:2003gv,Piekarewicz:2006ip,
Tsoneva:2007fk,Klimkiewicz:2007zz,Carbone:2010az}.  Recent pioneering
experiments on unstable neutron-rich isotopes in Sn, Sb, and Ni seem
to support this assertion~\cite{Adrich:2005,
Klimkiewicz:2007zz,Wieland:2009}.  

The main goals of this manuscript are twofold.  First, to use the
recently measured distribution of Pygmy dipole strength in
${}^{68}$Ni~\cite{Wieland:2009} to confirm our earlier assertion that
models with overly large neutron skins---and thus stiff symmetry
energies---are in conflict with experiment~\cite{Piekarewicz:2006ip}.
Second, to explore possible correlations between the neutron-skin
thickness in ${}^{208}$Pb and the {\sl dipole polarizability}. This is
motivated by a recent work by Reinhard and Nazarewicz that
suggests that {\sl the neutron skin is strongly correlated with the
dipole polarizability but very weakly correlated with the low-energy
electric dipole strength}~\cite{Reinhard:2010wz}.

Regarding the first goal, a significant first step was recently taken
by Carbone and collaborators~\cite{Carbone:2010az}. Using the fraction
of the EWSR exhausted by the Pygmy resonance in both
${}^{68}$Ni~\cite{Wieland:2009} and ${}^{132}$Sn~\cite{Adrich:2005},
the slope of the symmetry energy was constrained to the range
$L\!=\!64.8\pm15.7$~MeV. This constrain appears to rule out about half
of the 26 effective interactions---including all relativistic
ones---considered in their work. Viewed in this context, our
contribution is a modest one, as we will limit ourselves to a smaller
set of exclusively relativistic mean-field interactions. However, we
will show that some modern relativistic effective interactions are
soft enough to fall comfortably within the proposed range of $L$.
Concerning the second goal, the claim by Reinhard and
Nazarewicz~\cite{Reinhard:2010wz} is particularly intriguing given
that the dipole polarizability is proportional to the {\sl inverse
energy weighted sum} of the dipole response. As such, the dipole
polarizability weighs more heavily the low-energy (Pygmy) than the
high-energy (Giant) part of the response. So whereas the percentage of
the EWSR exhausted by the Pygmy amounts to a meager 5-8\%, its
contribution to the dipole polarizability can reach values as high as
20-25\%.  So how can the neutron-skin thickness of ${}^{208}$Pb be
strongly correlated to the dipole polarizability but weakly correlated
to the low-energy dipole strength? This is a question well worth
exploring. Note that in this work we will not address whether the PDR
is collective or not (for some recent reviews see
Refs.~\cite{Paar:2007bk,Paar:2010ww}). We believe that independent of
the nature of the mode, the emergence of low-energy dipole strength as
nuclei develop a neutron-rich skin is an incontrovertible fact.

The manuscript has been organized as follows. In
Sec.~\ref{sec:formalism} we introduce the formalism used in this work
paying special attention to the various moments of the dipole
response. In Sec.~\ref{sec:results} results are presented for the
distribution of dipole strength using relativistic effective
interactions that span a wide range of values for the neutron-skin
thickness of ${}^{208}$Pb.  We end by summarizing our results in
Sec.~\ref{sec:conclusions}.

\section{Formalism}
\label{sec:formalism}

The starting point for the calculation of the nuclear response is the
interacting Lagrangian density of Ref.~\cite{Mueller:1996pm}
supplemented by an isoscalar-isovector term originally introduced 
in Ref.~\cite{Horowitz:2000xj}. That is,
\begin{eqnarray}
{\mathscr L}_{\rm int} &=&
\bar\psi \left[g_{\rm s}\phi   \!-\! 
         \left(g_{\rm v}V_\mu  \!+\!
    \frac{g_{\rho}}{2}{\mbox{\boldmath $\tau$}}\cdot{\bf b}_{\mu} 
                               \!+\!    
    \frac{e}{2}(1\!+\!\tau_{3})A_{\mu}\right)\gamma^{\mu}
         \right]\psi \nonumber \\
                   &-& 
    \frac{\kappa}{3!} (g_{\rm s}\phi)^3 \!-\!
    \frac{\lambda}{4!}(g_{\rm s}\phi)^4 \!+\!
    \frac{\zeta}{4!}   g_{\rm v}^4(V_{\mu}V^\mu)^2 +
   \Lambda_{\rm v}\Big(g_{\rho}^{2}\,{\bf b}_{\mu}\cdot{\bf b}^{\mu}\Big)
                           \Big(g_{\rm v}^{2}V_{\nu}V^{\nu}\Big)\;.
 \label{LDensity}
\end{eqnarray}
The Lagrangian density includes an isodoublet nucleon field ($\psi$)
interacting via the exchange of two isoscalar mesons, a scalar
($\phi$) and a vector ($V^{\mu}$), one isovector meson ($b^{\mu}$),
and the photon ($A^{\mu}$)~\cite{Serot:1984ey,Serot:1997xg}. In
addition to meson-nucleon interactions the Lagrangian density is
supplemented by four nonlinear meson interactions with coupling
constants denoted by $\kappa$, $\lambda$, $\zeta$, and $\Lambda_{\rm
v}$.  The first two terms ($\kappa$ and $\lambda$) are responsible for
a softening of the equation of state of symmetric nuclear matter at
normal density~\cite{Boguta:1977xi}.
This softening results in a significant reduction of
the compression modulus of nuclear matter relative to the
original Walecka
model~\cite{Walecka:1974qa,Boguta:1977xi,Serot:1984ey} that is
demanded by the measured distribution of isoscalar
monopole strength in medium to heavy
nuclei~\cite{Youngblood:1999,Piekarewicz:2001nm,Piekarewicz:2002jd,
Colo:2004mj}. Further, omega-meson self-interactions, as described by
the parameter $\zeta$, also serve to soften the equation of state of
symmetric nuclear matter but at much higher densities.  Finally,
$\Lambda_{\rm v}$ induces isoscalar-isovector mixing and is
responsible for modifying the poorly-constrained density dependence of
the symmetry energy~\cite{Horowitz:2000xj,Horowitz:2001ya}.  Tuning
this parameter has served to uncover correlations between the neutron
radius of a heavy nucleus (such as ${}^{208}$Pb) and a host of both
laboratory and astrophysical observables.

The first step in a consistent mean-field plus RPA (MF+RPA) approach
to the nuclear response is the calculation of various ground-state
properties. This procedure is implemented by solving the equations of
motion associated with the above Lagrangian density in a
self-consistent, mean-field approximation~\cite{Serot:1984ey}.  For
the various meson fields one must solve Klein-Gordon equations with
the appropriate baryon densities appearing as their source
terms. These baryon densities are computed from the nucleon orbitals
that are, in turn, obtained from solving the one-body Dirac equation in
the presence of scalar and time-like vector potentials.  This
procedure must then be repeated until self-consistency is achieved.
What emerges from such a calculation is a set of single-particle
energies, a corresponding set of Dirac orbitals, and scalar and
time-like vector mean-field potentials. A detailed implementation of
this procedure may be found in Ref.~\cite{Todd:2003xs}.

Having computed various ground-state properties one is now in a
position to compute the linear response of the mean-field ground
state to a variety of probes.  In the present case we are interested 
in computing the electric dipole ($E1$) response as probed, for 
example, in photoabsorption experiments. Although the MF+RPA 
calculations presented here follow closely the formalism developed 
in much greater detail in Ref.~\cite{Piekarewicz:2001nm}, some 
essential details are repeated here for completeness.

The isovector dipole response of interest may be extracted from the 
imaginary part of a suitable polarization tensor. That is,
\begin{equation}
  S_{L}(q,\omega)=\sum_{n}\Big|\langle\Psi_{n}|\hat{\rho}({\bf q})|
  \Psi_{0}\rangle\Big|^{2}\delta(\omega-\omega_{n}) =-\frac{1}{\pi}
   \Im\,\Pi^{00}_{33}({\bf q},{\bf q};\omega) \;,
 \label{S03}
\end{equation}
where $\Psi_{0}$ is the exact nuclear ground state and $\Psi_{n}$ is 
an excited state with excitation energy $\omega_{n}\!=\!E_{n}\!-\!E_{0}$.
To excite electric-dipole modes a transition operator of the following
form is used:
\begin{equation}
 \hat{\rho}({\bf q}) \!=\! \int d^{3}r \, \bar{\psi}({\bf r}) 
  e^{-i{\bf q}\cdot{\bf r}} \gamma^{0}\tau_{3}\psi({\bf r}) \;.
 \label{Rhoq}
\end{equation}
Here $\hat{\rho}({\bf q})$ is the Fourier transform of the
timelike-isovector density, $\tau_{3}\!=\!{\rm diag}(1,-1)$ is the
third isospin matrix, and $\gamma^{0}\!=\!{\rm diag}(1,1,-1,-1)$ is the
zeroth (or timelike) component of the Dirac matrices.  Such a
transition operator is capable of exciting all natural-parity
states. In the present work we are interested in the study of electric
dipole modes so one must project out the $J^{\pi}\!=\!1^{-}$ component
of the transition operator. That is,
\begin{equation}
  \hat{\rho}_{\mu}({\bf q}) =-4\pi i Y_{1\mu}^{\ast}({\bf \hat{q}})
  \int d^{3}r \, \bar{\psi}({\bf r}) j_{1}(qr) Y_{1\mu}({\bf \hat{r}})
  \gamma^{0}\tau_{3}\psi({\bf r}) \;,
\end{equation}
where $j_{1}$ is a spherical Bessel function and $Y_{1\mu}$ are
spherical harmonics. To compare against experiment and to 
make contact with the model independent TRK sum rule we 
compute the response in the long-wavelength approximation.
Namely, we assume $j_{1}(qr)\!\approx\!qr/3$ so that the 
transition density reduces to
\begin{equation}
  \hat{\rho}_{\mu}({\bf q}) 
   \mathop{=}_{qr\ll 1}
   -\frac{4\pi}{3} i qY_{1\mu}^{\ast}({\bf \hat{q}})
   \int d^{3}r \, \bar{\psi}({\bf r}) rY_{1\mu}({\bf \hat{r}})
  \gamma^{0}\tau_{3}\psi({\bf r}) 
  \equiv -\frac{4\pi}{3}\imath qY_{1\mu}^{\ast}({\bf \hat{q}})
  {\mathscr M}(E1,\mu) \;,
\end{equation}
where ${\mathscr M}(E1,\mu)$ is the isovector-dipole
moment~\cite{Harakeh:2001}. In the long-wavelength limit, the
distribution of isovector dipole strength $R(\omega;E1)$ may be 
directly extracted from the longitudinal response. That is,
\begin{equation}
 \lim_{q\rightarrow 0} S_{L}(q,\omega;E1) =
 \frac{4\pi}{9}q^{2} R(\omega;E1)\;.
\end{equation}
where
\begin{equation}
 R(\omega;E1) = 3 \sum_{n} 
 \langle 1;n||{\mathscr M}(E1)||0\rangle^{2}
 \delta(\omega-\omega_{n}) =
 \sum_{n} B(E1;0\rightarrow n)
 \delta(\omega-\omega_{n}) \;.
\end{equation}

The cornerstone of the theoretical approach used here is the
polarization tensor introduced in Eq.~(\ref{S03}). For the
purposes of the present work, the relevant polarization tensor 
is defined in terms of a time-ordered product of two 
timelike-isovector densities [see Eq.~(\ref{Rhoq})]. That is,
\begin{equation}
  i\Pi^{00}_{33}(x,y) =  \langle \Psi_{0}| 
  T \Big[\hat{\rho}(x)\hat{\rho}(y)\Big] |\Psi_{0}\rangle 
   = \int_{-\infty}^{\infty} 
  \frac{d\omega}{2\pi}e^{-i\omega(x^{0}-y^{0})}
  \Pi^{00}_{33}({\bf x},{\bf y};\omega) \;.
\label{PiAlphaBeta}
\end{equation}
Connecting the nuclear response to the polarization tensor is highly
appealing as one can then bring to bear the full power of the
many-body formalism into the calculation of an experimental
observable~\cite{Fetter:1971}.  It is worth mentioning that the
polarization tensor contains all information about the excitation
spectrum of the system.  For example, in the uncorrelated case the
spectral content of the polarization tensor is both simple and
illuminating~\cite{Fetter:1971}.  The polarization tensor is an
analytic function of the excitation energy $\omega$---except for
simple poles located at $\omega\!=\!\epsilon_{p}\!-\!\epsilon_{h}$,
where $\epsilon_{p} (\epsilon_{h})$ are single-particle (single-hole)
energies. Moreover, the residues at these poles correspond to the
transition form-factors. Of course, selection rules enforce that only
particle-hole excitations with the correct quantum numbers can be
excited. To build collectivity into the nuclear response, all these
single-particle excitations must be correlated (or mixed) via the
residual particle-hole interaction. This is implemented by iterating
the uncorrelated polarization tensor to all orders.  Such a procedure
yields the RPA (Dyson's) equations whose solution embodies the
collective response of the ground state~\cite{Fetter:1971}. However,
the consistency of the MF+RPA approach hinges on the use of a residual
particle-hole interaction that is identical to the one used to
generate the mean-field potentials. Only then can one ensure the
preservation of important symmetries, such as the conservation of the
vector current and the decoupling of various spurious
modes~\cite{Dawson:1990wp,Piekarewicz:2001nm}. A more detailed
discussion of the relativistic MF+RPA approach may be found in
Ref.~\cite{Piekarewicz:2001nm} (see Ref.~\cite{Piekarewicz:2006ip} for
a discussion limited to the dipole response).  In the next
section results will be presented for the distribution of
dipole strength in ${}^{68}$Ni. Particular attention will be
placed on the various moments of the distribution and on their
relation to the corresponding moments of the photoabsorption cross
section. Hence, we close this section with a few essential definitions
and relations.

We start by introducing the photoabsorption cross section
$\sigma(\omega)$~\cite{Harakeh:2001}:
\begin{equation}
 \sigma(\omega) = \frac{4\pi^{2}e^{2}}{\hbar c}
 \sum_{n} \omega_{n} \langle 1;n||\hat{D}||0\rangle^{2}
 \delta(\omega-\omega_{n})  = \frac{16\pi^{3}}{9} 
 \frac{e^{2}}{\hbar c} \omega R(\omega) \;.
 \label{PhotoAbs}
\end{equation}
Here $\hat{D}$ is the standard dipole operator and note that the
``$E1$'' label has been suppressed from $R(\omega;E1)$. Having
established the connection between the dipole response and the
photoabsorption cross section we now proceed to compute their various
moments. These are defined as follows:
\begin{subequations}    
\begin{eqnarray}
 m_{n} &\equiv& \int_{0}^{\infty}\!\omega^{n} R(\omega) d\omega \;. \\
 \sigma_{n} &\equiv& \int_{0}^{\infty}\!\omega^{n} \sigma(\omega)
 d\omega  = \frac{16\pi^{3}}{9} \frac{e^{2}}{\hbar c} m_{n+1} \;.
\end{eqnarray}
\label{Moments}
\end{subequations}
In particular, the photoabsorption cross section satisfies the model 
independent TRK sum rule ($\sigma_{0}$) which is related to the EWSR 
(or $m_{1}$ moment) as follows:
\begin{subequations}
\begin{eqnarray}
 m_{1} &=&\sum_{n} \omega_{n} B(E1;0\rightarrow n) =
 \frac{9\hbar^{2}}{8\pi m}\left(\frac{NZ}{A}\right) \approx
 14.8 \left(\frac{NZ}{A}\right) {\rm fm}^{2}{\rm MeV} \;, \\
 \sigma_{0} &=& \frac{16\pi^{3}}{9} \frac{e^{2}}{\hbar c} m_{1} =
 2\pi^{2} \frac{e^{2}}{\hbar c} \frac{(\hbar c)^{2}}{mc^{2}} 
 \left(\frac{NZ}{A}\right) 
 \approx 60 \left(\frac{NZ}{A}\right) {\rm MeV \, mb}\;.
\end{eqnarray}
\label{TRK}
\end{subequations}
In addition to the fundamental TRK sum rule, a moment of critical
importance to the present work because of its sensitivity to the
symmetry energy is the {\sl dipole polarizability}. The dipole
polarizability is particularly sensitive to low-energy dipole strength
given that it is directly proportional to the {\sl inverse} energy
weighted sum $m_{-1}$ (or $\sigma_{-2}$). That is,
\begin{equation}
  \alpha_{D} = 2e^{2} \sum_{n} 
  \frac{\langle 1;n||\hat{D}||0\rangle^{2}}{\omega_{n}} =
  \frac{\hbar c}{2\pi^{2}}\sigma_{-2}=
  \frac{8\pi}{9}e^{2}m_{-1}\;.
\label{DipPol}
\end{equation}

\section{Results}
\label{sec:results}

\begin{figure}[ht]
\vspace{-0.05in}
\includegraphics[width=0.6\columnwidth,angle=0]{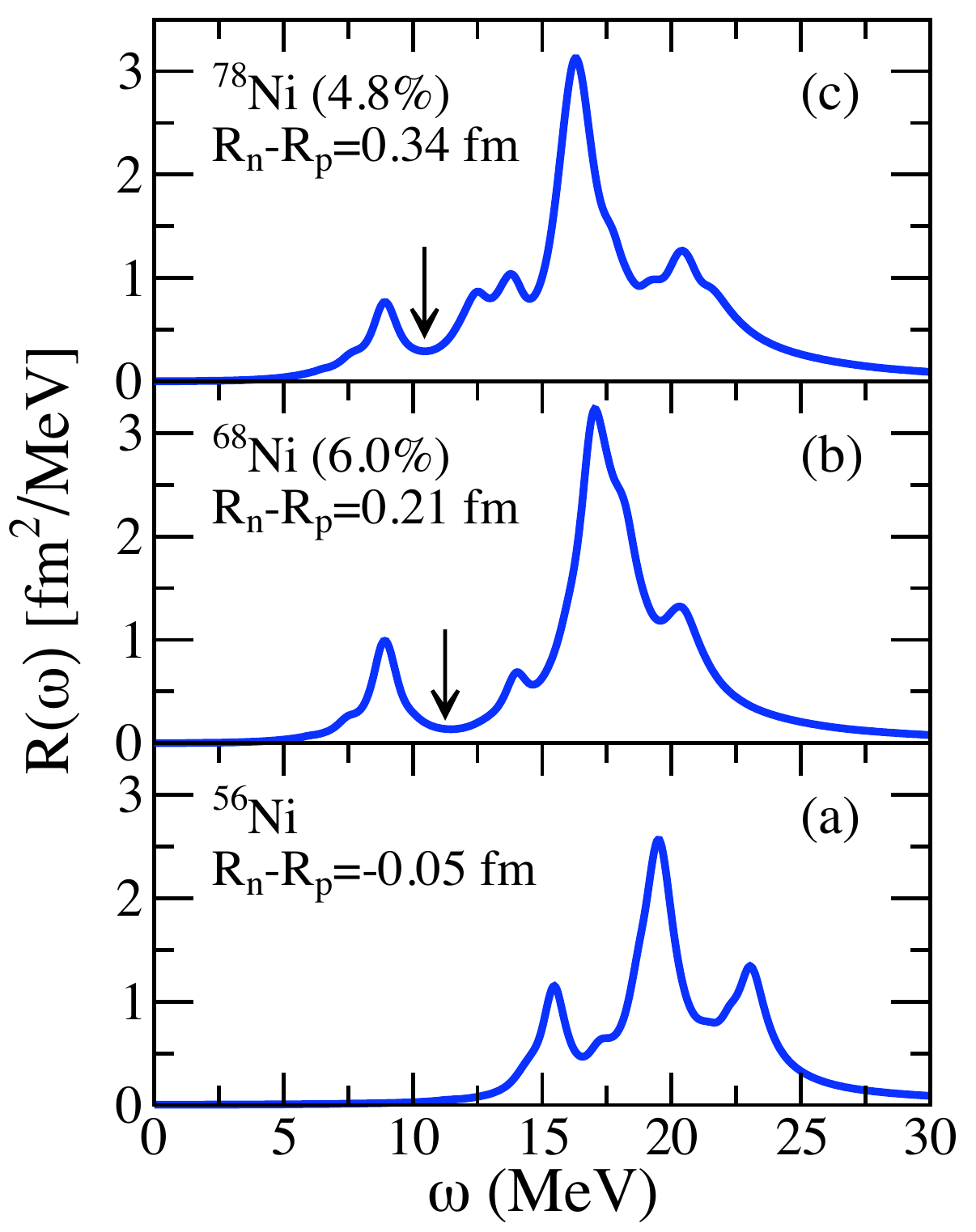}
\caption{Distribution of isovector dipole strength for the three
              closed-(sub)shell nickel isotopes: ${}^{56}$Ni,
              ${}^{68}$Ni, and ${}^{78}$Ni. Mean field plus RPA
              predictions are shown using the FSUGold parameter 
              set~\cite{Todd-Rutel:2005fa}.}
\label{Fig1}
\end{figure}

We start this section by displaying in Fig.~\ref{Fig1} the
distribution of dipole strength for the three closed-shell (or at
least closed-subshell) nickel isotopes ${}^{56}$Ni, ${}^{68}$Ni, and
${}^{78}$Ni. Predictions are displayed using the accurately calibrated
FSUGold (or ``FSU'' for short)
parametrization~\cite{Todd-Rutel:2005fa}.  In order to resolve
individual transitions to bound single-particle states, a small
artificial width of $0.5$~MeV was included in the calculations.  Note,
however, that because the {\sl non-spectral} character of our RPA
approach~\cite{Piekarewicz:2001nm}, particle-escape widths are
computed exactly within the model. In panel (a) we display the
distribution of dipole strength for the doubly-magic nucleus
${}^{56}$Ni. Predictions for the neutron skin of this $N\!=\!Z$
nucleus yield a small negative value of $R_{n}\!-\!R_{p}\!=\!-0.05$~fm
because of the Coulomb repulsion among the protons. Consistent with
the notion that the Pygmy dipole resonance represents an oscillation
of the neutron-rich skin against the symmetric core, no low-energy
dipole strength is found. Instead, all the dipole strength is found in
the region of the Giant resonance (at excitation energies
$\gtrsim\!12$~MeV) and exhausts 114\% of the EWSR.  In
Fig.~\ref{Fig1}(b) we observe a significant qualitative change as one
moves from ${}^{56}$Ni to the neutron-rich ${}^{68}$Ni isotope.  In a
mean-field approach such as the one adopted here, the extra 12
neutrons fill in the $1f^{5/2}$, $2p^{3/2}$, and $2p^{1/2}$
orbitals. This leads to the development of a fairly large neutron-skin
thickness of $R_{n}\!-\!R_{p}\!=\!0.21$~fm. Strongly correlated to the
development of the neutron skin is the appearance of low-energy dipole
strength. Indeed, we now find that 6\% of EWSR is contained in the
low-energy region. Note that a {\sl model dependent} choice must
be made on how to separate the low- and high-energy regions. In the
present model---indeed in most MF+RPA models---this separation is
natural. Here we define $\omega_{t}\!\equiv\!11.25$~MeV as the
``Pygmy-to-Giant'' transition energy [see arrow on
Fig.~\ref{Fig1}(b).]  Finally, we display in Fig.~\ref{Fig1}(c) the
distribution of dipole strength for the very exotic ${}^{78}$Ni
nucleus. Whereas the additional 10 neutrons filling the $1g^{9/2}$
orbital contribute to a further increase in the thickness of its
neutron skin (from 0.21 to 0.34~fm) the fraction of the EWSR exhausted
by the Pygmy resonance actually goes down; from 6\% to 4.8\%. This
{\sl anti-correlation} is reminiscent of the one we reported in Fig.~4
of Ref.~\cite{Piekarewicz:2006ip} for the case of the Tin-isotopes. In
the present case, as the {\sl ``intruder''} $1g^{9/2}$ orbital is
pulled down by the strong spin-orbit force, we observe a mild
enhancement of the total EWSR.  This enhancement is consistent with
the few percent increase in the $NZ/A$ factor as one goes from
${}^{68}$Ni to ${}^{78}$Ni [see Eq.~(\ref{TRK})].  However, such an
increase is concentrated in the region of the giant resonance because
dipole excitations originating in the $1g^{9/2}$ orbital (and ending
in the $1h^{11/2}$, $1h^{9/2}$, and $2f^{7/2}$ orbitals) lie high in
energy.  Thus, we conclude that ${}^{68}$Ni is as good---indeed
better---than ${}^{78}$Ni in the search for correlations between the
neutron skin and low-energy dipole strength.  The same conclusion
applies for the case of the Tin isotopes: the {\sl stable}
${}^{120}$Sn isotope may be as useful as 
${}^{132}$Sn~\cite{Piekarewicz:2006ip}.

To search for correlations between the neutron-skin thickness of
${}^{208}$Pb and the development of Pygmy strength in ${}^{68}$Ni we
introduce---in addition to FSUGold~\cite{Todd-Rutel:2005fa}---the
accurately-calibrated NL3 effective
interaction~\cite{Lalazissis:1996rd,Lalazissis:1999}. Parameter sets
for these two models are listed in Table~\ref{Table1}.  Although
enormously successful in reproducing ground-state energies and charge
radii, NL3 predicts equations of state for both symmetric and pure
neutron matter that appear too stiff when compared against theoretical
and experimental
constraints~\cite{Danielewicz:2002pu,Piekarewicz:2003br,Piekarewicz:2009gb}.
To remedy this deficiency the FSUGold parameter set includes two
additional empirical constants, namely, $\zeta$ and $\Lambda_{\rm v}$
[see Eq.~(\ref{LDensity})].  To fully explore the sensitivity of the
low energy dipole strength to changes in the neutron-skin thickness of
${}^{208}$Pb we introduce a {\sl ``family''} of NL3 and FSUGold
models.  These families are generated by following a procedure first
introduced in Ref.~\cite{Horowitz:2000xj}. This procedure is
implemented by changing the isovector parameters $\Lambda_{\rm v}$ and
$g_{\rho}$ in such a way that the value of the symmetry energy remains
fixed at $\approx\!26$~MeV at a baryon density of $\approx\!0.1~{\rm
fm}^{-3}$.  This prescription ensures that well constrained
observables---such as masses and charge radii---remain consistent with
their experimental values.  We display in Table~\ref{Table2} the
appropriate isovector parameters $\Lambda_{\rm v}$ and $g_{\rho}$ for
the NL3 and FSUGold families of mean-field interactions. Together with
those parameters we display two observables that are particularly
sensitive to these changes, namely, the slope of the symmetry energy
at saturation density ($L$) and the neutron skin thickness
($R_{n}$-$R_{p}$) of both ${}^{208}$Pb and ${}^{68}$Ni (with the latter 
shown in parenthesis).

\begin{widetext}
\begin{center}
\begin{table}[h]
\begin{tabular}{|l||c|c|c|c|c|c|c|c|c|c|}
 \hline
 Model & $m_{\rm s}$  & $m_{\rm v}$  & $m_{\rho}$  
       & $g_{\rm s}^2$ & $g_{\rm v}^2$ & $g_{\rho}^2$
       & $\kappa$ & $\lambda$ & $\zeta$ & $\Lambda_{\rm v}$\\
 \hline
 \hline
 NL3       & 508.194 & 782.501 & 763.000 & 104.3871 & 165.5854 &  79.6000 
               & 3.8599  & $-$0.015905 & 0.00 & 0.000 \\
 FSU        & 491.500 & 782.500 & 763.000 & 112.1996 & 204.5469 & 138.4701 
               & 1.4203  & $+$0.023762 & 0.06 & 0.030 \\
\hline
\end{tabular}
\caption{Parameter sets for the two accurately calibrated relativistic
               mean-field models used in the text:
               NL3~\cite{Lalazissis:1996rd,Lalazissis:1999}
               and FSUGold~\cite{Todd-Rutel:2005fa}. The parameter
               $\kappa$ and the meson masses $m_{\rm s}$, $m_{\rm v}$, 
               and $m_{\rho}$ are all given in MeV. The nucleon mass
               has been fixed at  $M\!=\!939$~MeV in both models.}
\label{Table1}
\end{table}
\end{center}
\end{widetext}

\begin{widetext}
\begin{center}
\begin{table}
\begin{tabular}{|l||c|c|c|c|}
  \hline
   Model &  $\Lambda_{\rm v}$ & $g_{\rho}^2$ & $L$~(MeV) 
             &  $R_{n}$-$R_{p}$~(fm) \\   
  \hline
  \hline
   NL3    
   &  0.00 &  79.6000  & 118.189 & 0.280~(0.261) \\
   &  0.01 &  90.9000  &  87.738  & 0.251~(0.241) \\
   &  0.02 & 106.0000 &  68.217  & 0.223~(0.222) \\
   &  0.03 & 127.1000 &  55.311  & 0.195~(0.203) \\
   &  0.04 & 158.6000 &  46.607  & 0.166~(0.183) \\
 \hline   
 FSU
   &  0.00 &  80.2618  & 108.764 & 0.286~(0.265) \\    
   &  0.01 &  93.3409  &  87.276  & 0.260~(0.248) \\
   &  0.02 & 111.5126 &  71.833  & 0.235~(0.223) \\
   &  0.03 & 138.4701 &  60.515  & 0.207~(0.211) \\
   &  0.04 & 182.6162 &  52.091  & 0.176~(0.189) \\
\hline
\end{tabular}
 \caption{The NL3 and FSUGold {\sl ``families''} of mean-field interactions.
 The isovector parameters $\Lambda_{\rm v}$ and $g_{\rho}$ were adjusted so
 that all models have the same symmetry energy of $\approx\!26$~MeV at
 a density of $\approx\!0.1~{\rm fm}^{-3}$.
 Tuning $\Lambda_{\rm v}$ significantly affects the slope of the
 symmetry energy at saturation density $L$ and the neutron-skin thickness 
 of ${}^{208}$Pb and ${}^{68}$Ni (the latter shown in parenthesis).}
\label{Table2}
\end{table}
\end{center}
\end{widetext}

\begin{figure}[h]
\vspace{-0.05in}
\includegraphics[width=0.4\columnwidth,angle=0]{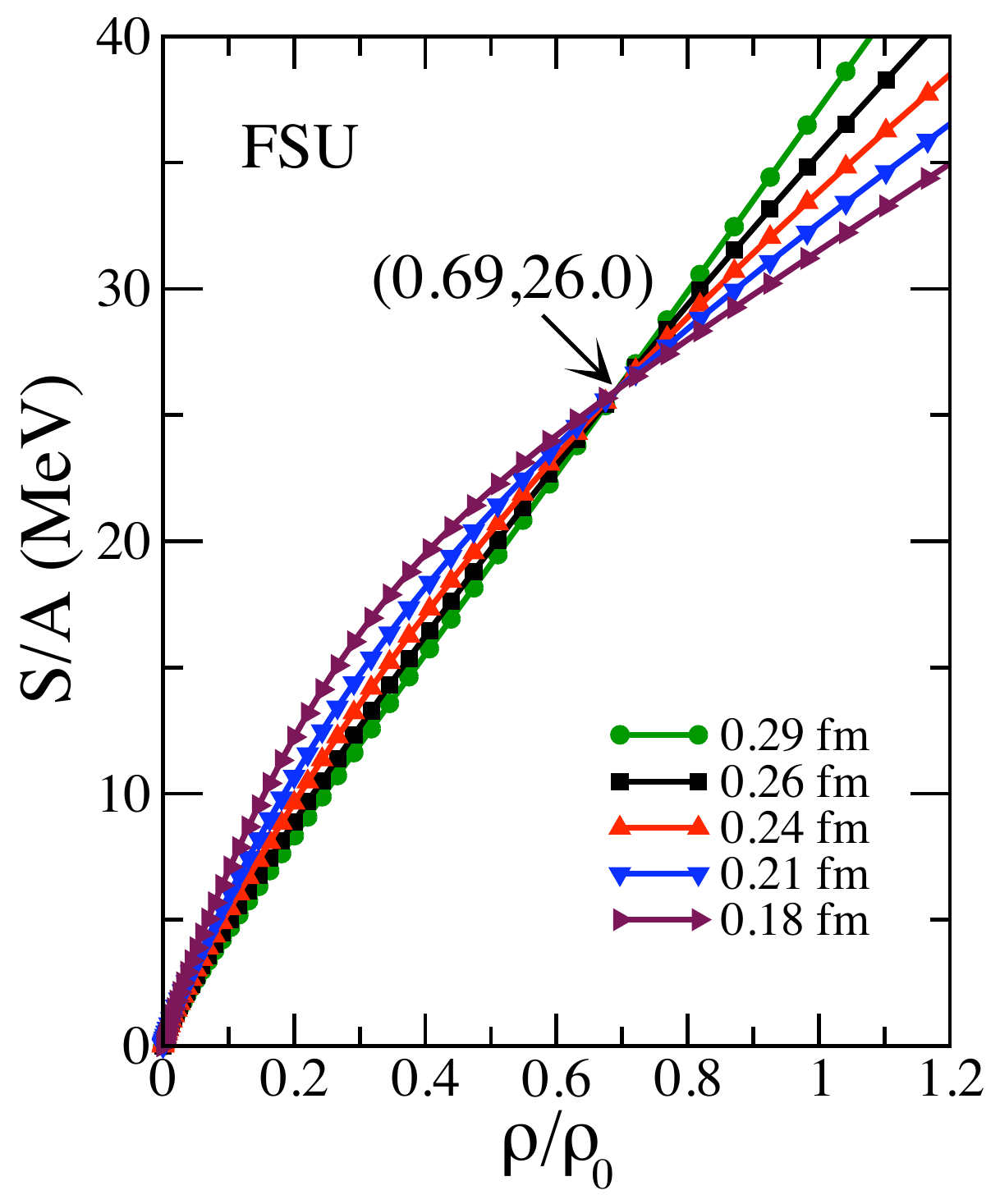}
\caption{(Color online) Symmetry energy per nucleon as a function of 
 density (in units of the saturation density). The various effective 
 interactions are labeled according to their predictions for the 
 neutron-skin thickness of ${}^{208}$Pb.}
\label{Fig2}
\end{figure}

To illustrate the role of the various mean-field interactions
we display in Fig.~\ref{Fig2} the symmetry energy per nucleon
predicted by all these models. The density dependence of the symmetry
energy is of central importance to this work as it will be used to
correlate the neutron-skin thickness of ${}^{208}$Pb to the dipole
strength in ${}^{68}$Ni. The symmetry energy represents the
susceptibility of the system to changes in the neutron-proton
asymmetry. It is defined as follows:
\begin{equation}
  S(\rho) = \frac{1}{2}\left( \frac{\partial^{2}E(\rho,\alpha)}
  {\partial \alpha^{2}}\right)_{\alpha=0} 
  \approx E_{\rm PNM}(\rho) -  E_{\rm SNM}(\rho) \;,
 \label{SymE}
\end{equation}
where $\rho\!=\!\rho_{n}\!+\!\rho_{p}$ is the baryon density of the
system and $\alpha\!=\!(\rho_{n}\!-\!\rho_{p})/\rho$ the
neutron-proton asymmetry. As indicated in Eq.~(\ref{SymE}), the
symmetry energy is to a very good approximation equal to the energy of
pure neutron matter minus that of symmetric matter. In Fig.~\ref{Fig2}
the convergence of all models at a density of $\approx\!0.1~{\rm
fm}^{-3}$ (or $\approx\!0.7$ times the density of nuclear matter at
saturation) is clearly discernible. However, the departure from this
common value is model dependent. For example, models with a {\sl
stiff} symmetry energy---namely, those that change rapidly with
density---predict a large slope at saturation density and a
correspondingly large value for the neutron-skin thickness of
${}^{208}$Pb. Note that $L$ is proportional to the pressure of pure
neutron matter; hence, the larger the value of the neutron pressure
the larger the neutron skin.  Also note that models with a stiff
symmetry energy predict a {\sl lower symmetry energy} at low
densities as compared to their softer counterparts. Given that the
symmetry energy ($S\!\propto\!\partial^{2}E/\partial\alpha^{2}$) acts
as the restoring force for isovector modes, we expect that as the
symmetry energy stiffens, the distribution of isovector dipole
strength will become softer. These arguments suggest how to exploit
the behavior of the symmetry energy to correlate the neutron-skin
thickness of ${}^{208}$Pb to the dipole strength in
${}^{68}$Ni.

\begin{figure}[h]
\vspace{-0.05in}
\includegraphics[width=0.45\columnwidth,angle=0]{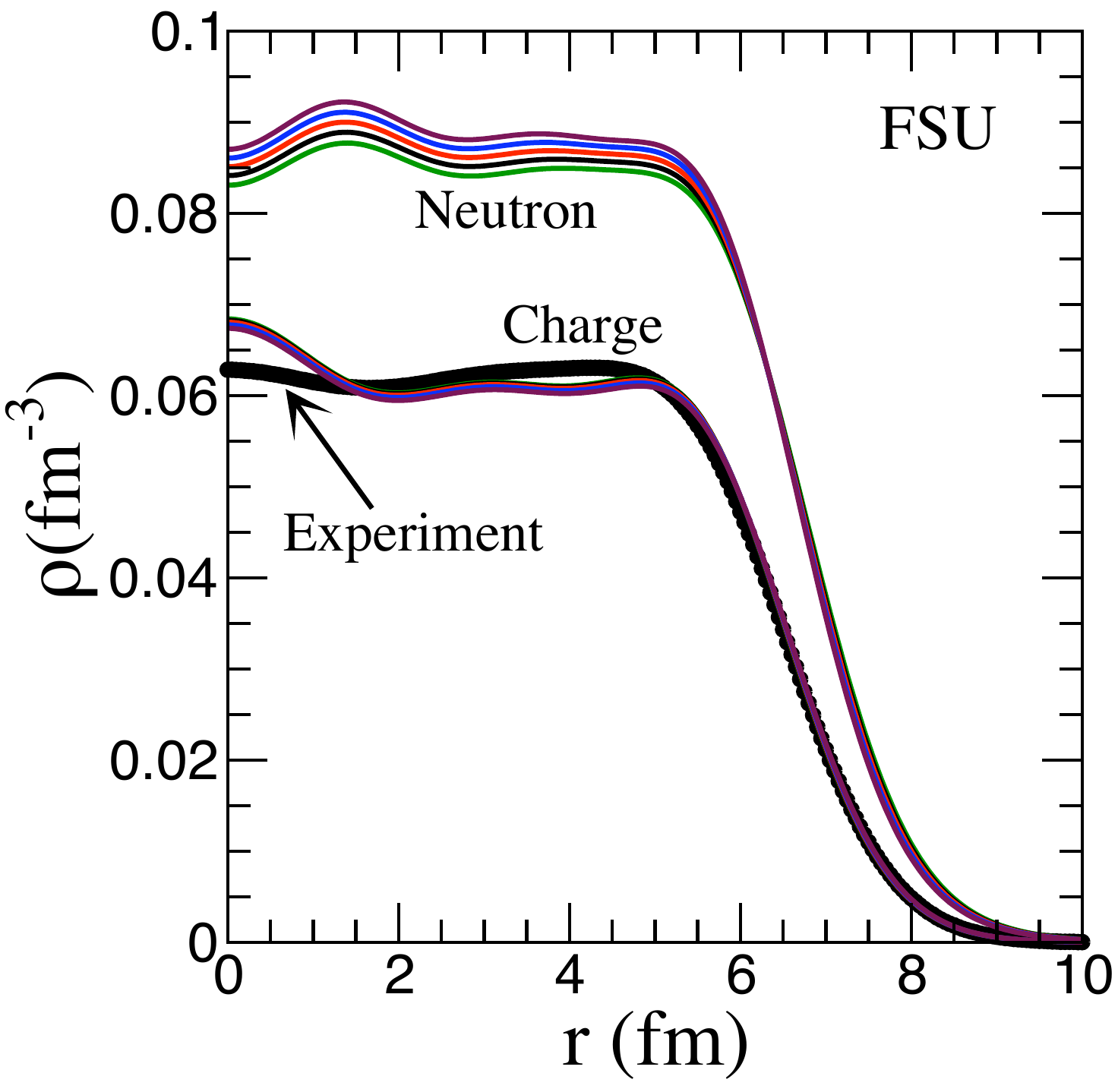}
\caption{(Color online) Model predictions for the charge 
               and neutron densities of ${}^{208}$Pb using the FSU
               family of effective interactions. The experimental 
               charge density is from Ref.~\cite{DeJager:1987qc}.}
\label{Fig3}
\end{figure}

Before doing so, however, we further validate the prescription used to
generate the FSU family of mean-field interactions by displaying in
Fig.~\ref{Fig3} charge and neutron densities for ${}^{208}$Pb (similar
results are obtained in the case of NL3). Whereas significant
differences are easily discerned in the predictions of the (unknown)
neutron density, the model dependence is very small in the case of the
charge density. For example, all models predict a mean-square charge
radius that is within 0.4\% of the experimental value. In the case of
the binding energy of ${}^{208}$Pb, the agreement with experiment is
even better (by about one order of magnitude).  Yet a fairly simple
modification to the isovector interaction allows one to generate a
fairly wide range of values for the neutron-skin thickness of
${}^{208}$Pb: from 0.18 to 0.29~fm.  

\begin{figure}[h]
\vspace{-0.05in}
\includegraphics[width=0.48\columnwidth,angle=0]{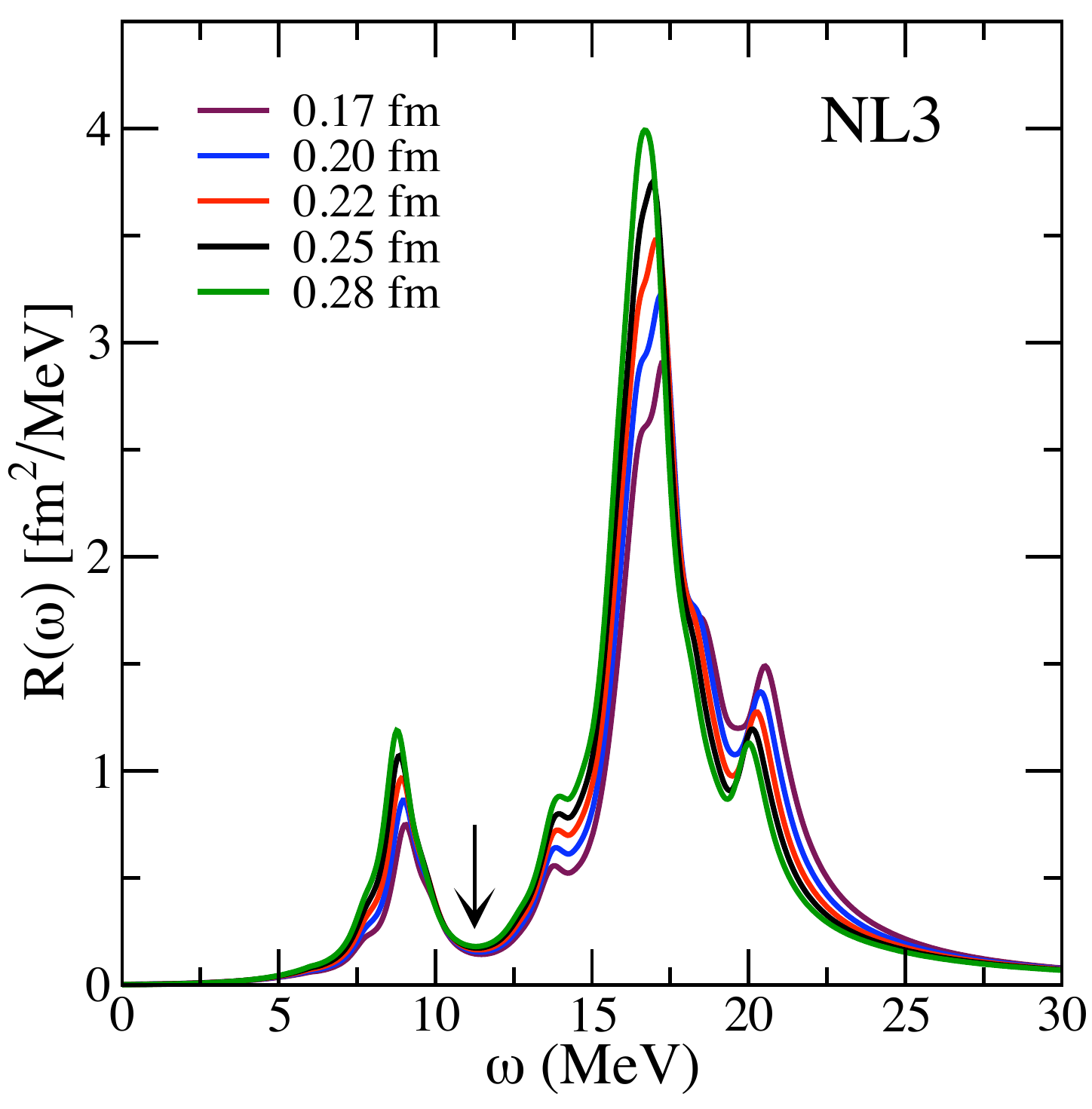}
\includegraphics[width=0.48\columnwidth,angle=0]{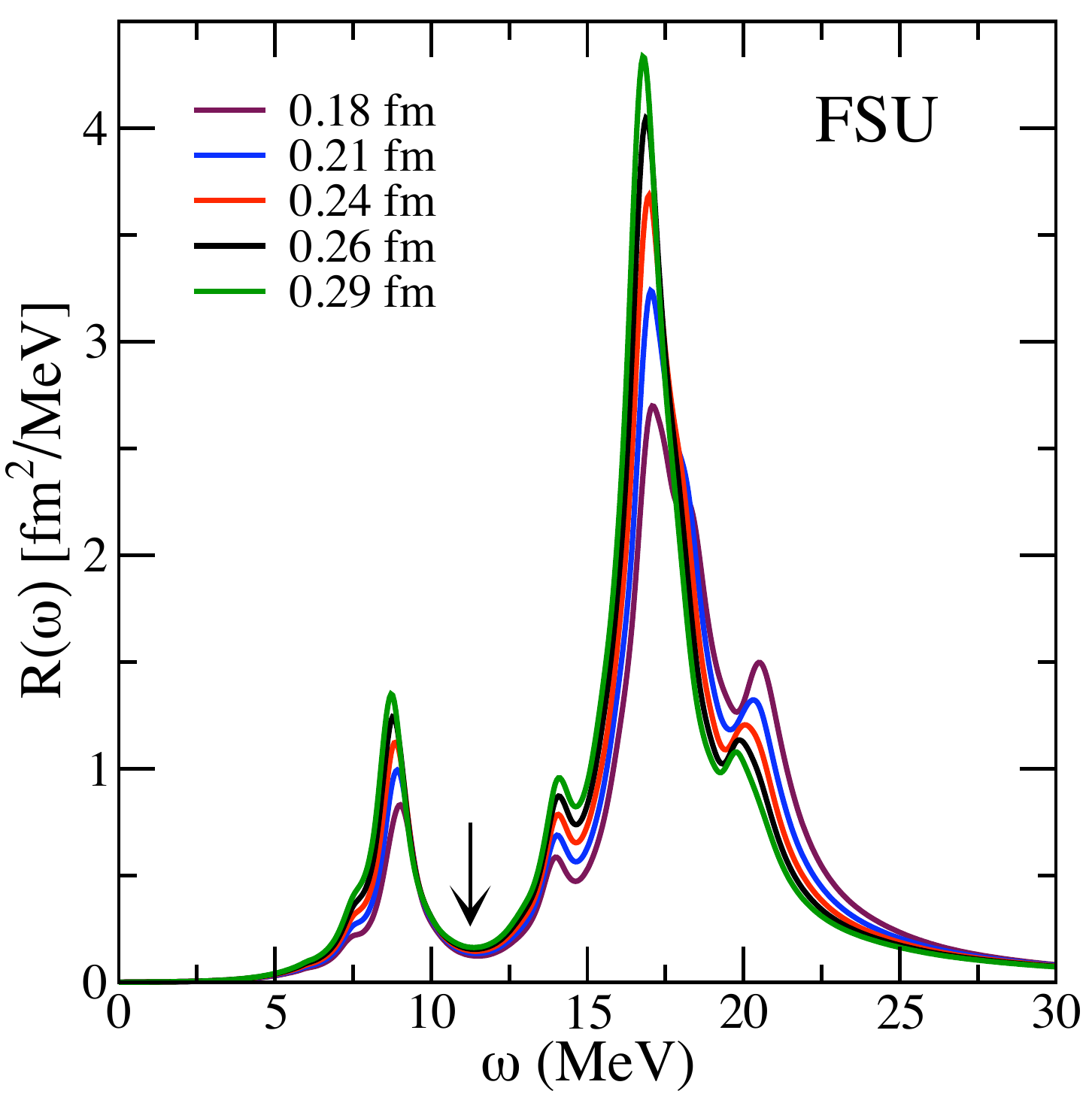}
\caption{(Color online) Distribution of dipole strength
 in ${}^{68}$Ni computed in a MF+RPA approach using the NL3
 (left panel) and FSU (right panel) families of effective interactions.}
\label{Fig4}
\end{figure}

Having validated the choice of mean-field models employed in this
work, we now proceed to display in Fig.~\ref{Fig4} their predictions
for the distribution of dipole strength in ${}^{68}$Ni. Various
moments of the distribution as well as the TRK enhancement factor
$\kappa_{TRK}$~\cite{Harakeh:2001} have also been collected in
Table~\ref{Table3}.  Given that the $m_{-1}$ moment is simply related
to the dipole polarizability $\alpha_{D}$ [see Eq.~(\ref{DipPol})], it is the
latter that is listed in Tables~\ref{Table3}.  Also note that the
various curves have been labeled according to their prediction for the
neutron-skin thickness of ${}^{208}$Pb.  The distribution of strength
naturally separates into low-energy (Pygmy) and high-energy (Giant)
regions. To compute the contribution from these two regions to the
various moments we have selected the Pygmy-to-Giant transition energy
to be equal to $\omega_{t}\!\equiv\!11.25$~MeV, as indicated in the
figure.  As argued earlier, models with a soft symmetry energy predict
large symmetry energies at the low densities of relevance to the
excitation of the dipole mode (see Fig.~\ref{Fig2}). In turn, such a
large restoring force generates a significant {\sl hardening and
quenching} of the response. That is, models with a soft symmetry
predict a distribution of strength that is both hardened ({\sl i.e.,}
pushed to higher excitation energies) and quenched relative to their
stiffer counterparts. These facts are readily discernible in
Fig.~\ref{Fig4}.  We now proceed to explore the consequences of such
a hardening and quenching on the various moments of the dipole
response.

\begin{center}
\begin{table}
\begin{tabular}{|l||c|c|c|c|c|}
  \hline
   Model & $\Lambda_{\rm v}$ & $m_{1}$~(fm${}^{2}$MeV) 
   & $\kappa_{\rm TRK}$ & $m_{0}$~(fm${}^{2}$) 
   & $\alpha_{D}$~(fm${}^{3}$) \\ 
 \hline
  \hline
   NL3    
   & 0.00 & 284.548 & 0.163 & 17.428 & 4.716 \\
   & 0.01 & 284.190 & 0.162 & 17.151 & 4.570 \\
   & 0.02 & 283.320 & 0.158 & 16.832 & 4.412 \\
   & 0.03 & 281.948 & 0.153 & 16.454 & 4.235 \\
   & 0.04 & 279.749 & 0.144 & 15.984 & 4.026 \\
  \hline
   FSU    
   & 0.00 & 283.364 & 0.158 & 17.270 & 4.664 \\
   & 0.01 & 282.667 & 0.156 & 16.976 & 4.511 \\
   & 0.02 & 281.535 & 0.151 & 16.627 & 4.339 \\
   & 0.03 & 279.784 & 0.144 & 16.199 & 4.138 \\
   & 0.04 & 276.857 & 0.132 & 15.635 & 3.887 \\
\hline
\end{tabular}
\caption{Various moments of the distribution of dipole
 strength $R(\omega)$ for the two families of relativistic mean-field 
 interactions defined in the text. Note that $\alpha_{D}$ is the
 dipole polarizability and $\kappa_{\rm TRK}$ denotes the 
 enhancement factor of the TRK sum rule.}
\label{Table3}
\end{table}
\end{center}

\begin{figure}[h]
\vspace{-0.05in}
\includegraphics[width=0.48\columnwidth,angle=0]{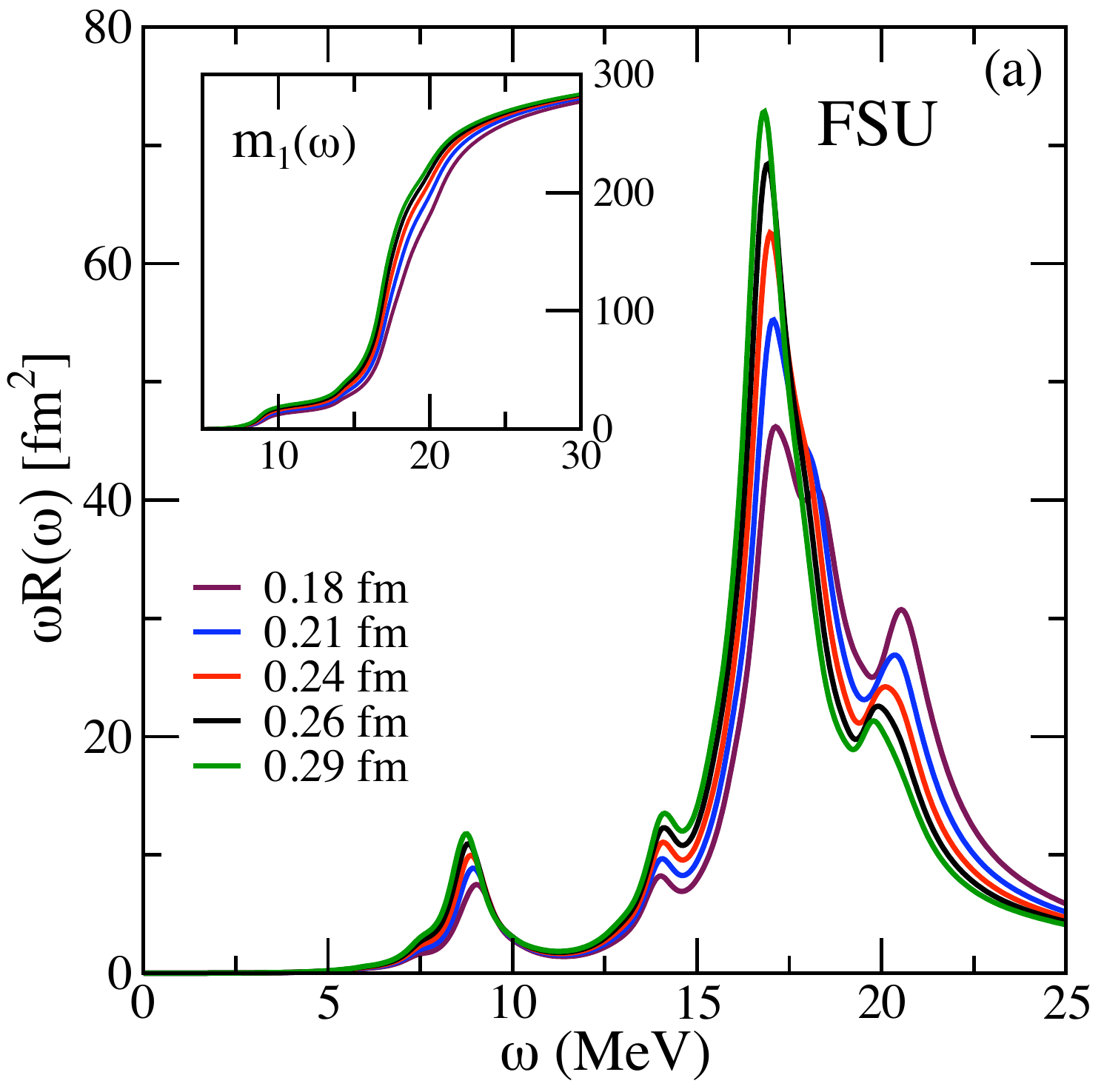}
\includegraphics[width=0.48\columnwidth,angle=0]{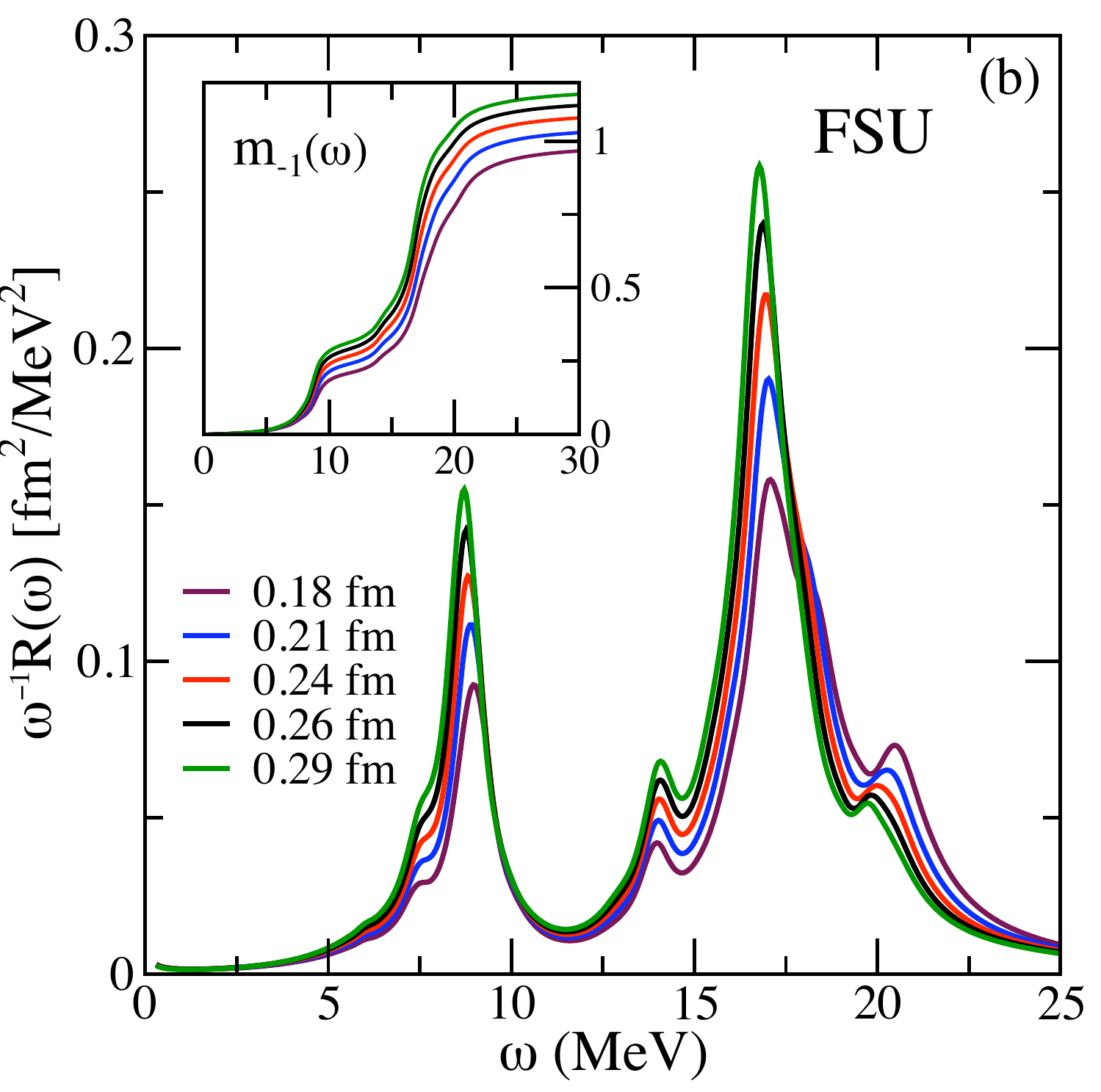}
\caption{(Color online) The energy weighted dipole 
 response (a) and the {\sl inverse} energy weighted dipole 
 response (b) in ${}^{68}$Ni computed with the FSU family of effective
 interactions. The insets display the cumulative sums as defined in
 Eqs.~(\ref{m1Def}) and~(\ref{mm1Def}).}
\label{Fig5}
\end{figure}

Although the previous discussion suggests a correlation between the
neutron-skin thickness of ${}^{208}$Pb and the distribution of dipole
strength, the alluded correlation may or may not extend to the various
moments of the distribution. This depends critically on whether the
quenching and hardening work for or against each other. A prototypical
case in which they work against each other is the total $m_{1}$
moment, {\sl i.e.,} the energy weighted sum. This must be so because
the $m_{1}$ moment satisfies a {\sl model-independent sum rule}. That
is, the energy weighting enhances the hardened response as to exactly
compensate for its original quenching. To appreciate this fact we
display in Fig.~\ref{Fig5}(a) the energy weighted dipole response
$\omega R(\omega)$.  Plotted in the inset is the {\sl cumulative}
contribution of $\omega R(\omega)$ to the EWSR defined as
\begin{equation}
   m_{1}(\omega) = \int_{0}^{\omega} \omega' R(\omega') d\omega'\;.
 \label{m1Def}
\end{equation}
The cumulative sum $m_{1}(\omega)$ displays what appears to be a mild
model dependence as it starts to accumulate strength in the region of
the Pygmy resonance. The largest model dependence develops around the
main giant-resonance peak. Yet any residual model dependence rapidly
disappears as the sum rule is exhausted---as it should be. However,
given that no sum rule protects the {\sl fraction} of the EWSR
contained in the low-energy region, a model dependence remains. This
generates a correlation between the neutron-skin thickness in
${}^{208}$Pb and the fraction of the EWSR exhausted by the Pygmy
resonance (see Table~\ref{Table4}).

\begin{center}
\begin{table}
\begin{tabular}{|l||c|c|c|c|c|}
  \hline
   Model & $\Lambda_{\rm v}$ & $E_{c}$~(MeV) &
   $m_{1}$~(fm${}^{2}$MeV) & $m_{0}$~(fm${}^{2}$) & 
   $\alpha_{D}$~(fm${}^{3}$) \\ 
 \hline
  \hline
   NL3    
   & 0.00 & 8.602 & 20.680~(7.268) & 2.404~(13.794) & 1.185~(25.133) \\
   & 0.01 & 8.643 & 19.231~(6.767) & 2.225~(12.973) & 1.099~(24.040) \\
   & 0.02 & 8.675 & 17.713~(6.252) & 2.042~(12.130) & 1.011~(22.907) \\
   & 0.03 & 8.700 & 16.074~(5.701) & 1.848~(11.229) & 0.919~(21.698) \\
   & 0.04 & 8.715 & 14.248~(5.093) & 1.635~(10.227) & 0.819~(20.344) \\
  \hline
   FSU    
   & 0.00 & 8.525 & 21.592~(7.620) & 2.533~(14.666) & 1.241~(26.601) \\
   & 0.01 & 8.565 & 20.185~(7.141) & 2.357~(13.883) & 1.151~(25.521) \\
   & 0.02 & 8.603 & 18.628~(6.617) & 2.165~(13.023) & 1.055~(24.322) \\
   & 0.03 & 8.640 & 16.871~(6.030) & 1.953~(12.054) & 0.950~(22.957) \\
   & 0.04 & 8.676 & 14.732~(5.321) & 1.698~(10.861) & 0.825~(21.234) \\
\hline
\end{tabular}
\caption{Contribution from the Pygmy-resonance region 
 ($0\!\le\!\omega\!\le11.25$~MeV) to the various moments of 
 the distribution of dipole strength. The centroid energy 
 has been defined as $E_{c}\!=\!m_{1}/m_{0}$ and the quantities in
 parenthesis denote the fraction of the total moment contained in 
 the region of the Pygmy resonance.}
\label{Table4}
\end{table}
\end{center}

\begin{figure}[h]
\vspace{-0.05in}
\includegraphics[width=0.49\columnwidth,angle=0]{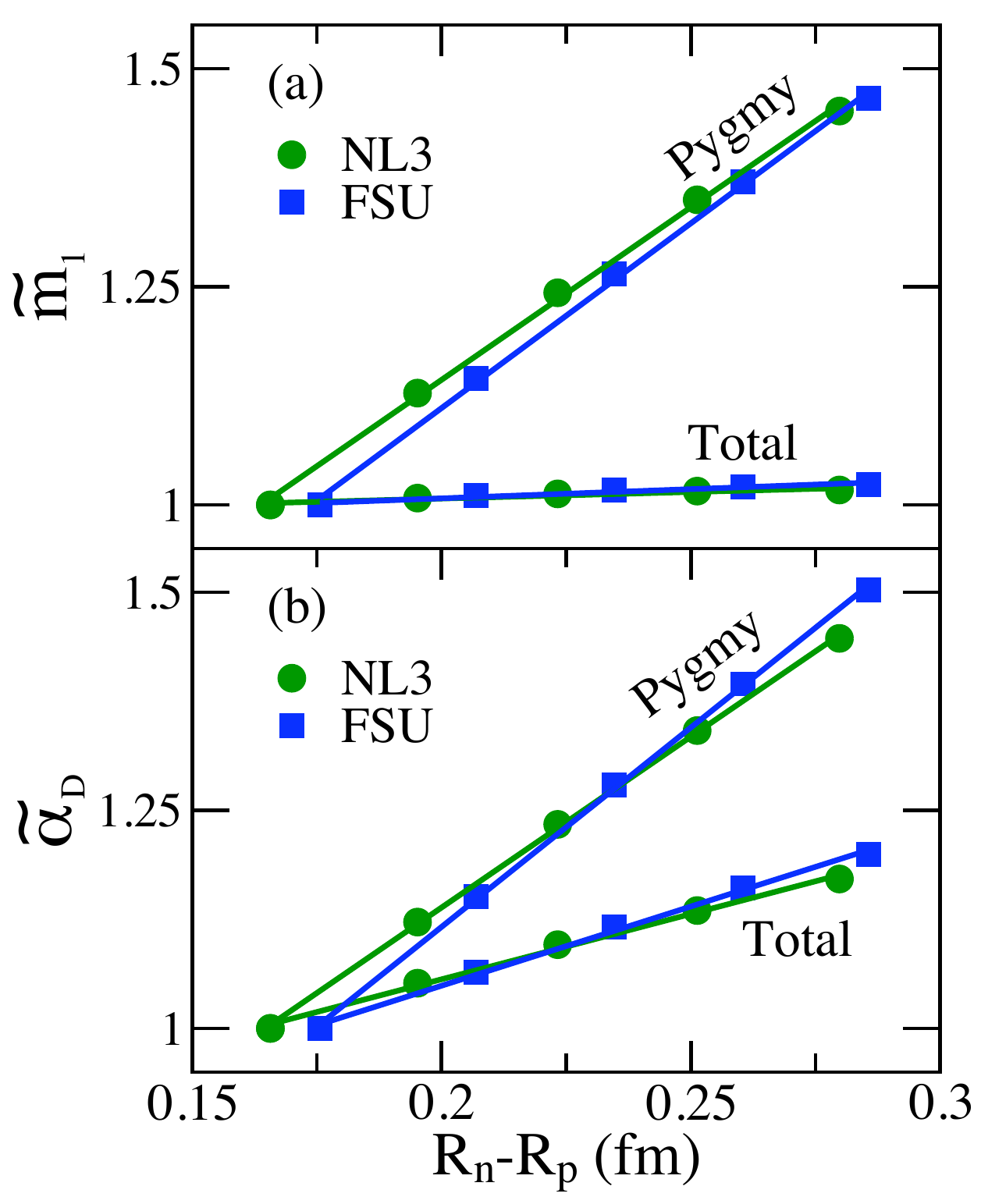}
\caption{(Color online)  {\sl Fractional change} in the energy 
 weighted sum rule (a) and dipole polarizability (b) for ${}^{68}$Ni 
 as a function of the neutron-skin thickness of ${}^{208}$Pb.}
\label{Fig6}
\end{figure}

A particularly attractive case in which both the hardening and the
quenching work in tandem is the {\sl inverse} energy weighted response
$\omega^{-1} R(\omega)$ displayed in Fig.~\ref{Fig5}(b). First, given
that the $\omega^{-1}$ factor enhances preferentially the low-energy
part of response, the Pygmy resonance now accounts for a significant
fraction---of about 20-25\%---of the $m_{-1}$ moment. This should be
contrasted against the EWSR where the Pygmy resonance exhausts merely
5-8\% of the sum rule; see Tables~\ref{Table3}-\ref{Table5}. Second,
the inverse energy weighting enhances the softer responses even
further.  Pictorially, this behavior is best illustrated in the inset
of Fig.~\ref{Fig5}(b) which displays the {\sl cumulative}
$m_{-1}(\omega)$ sum:
\begin{equation}
   m_{-1}(\omega) = \int_{0}^{\omega}\frac{R(\omega')}{\omega'} d\omega'\;.
 \label{mm1Def}
\end{equation}
The inset provides a clear indication that both the total $m_{-1}$
moment as well as the fraction contained in the Pygmy resonance are
highly sensitive to the neutron-skin thickness of ${}^{208}$Pb.  To
heighten this sensitivity we display in Fig.~\ref{Fig6} the {\sl
fractional change} in both the total and Pygmy contributions to the
$m_{1}$ moment and to the dipole polarizability $\alpha_{D}$ (we
denote these fractional changes with a {\sl ``tilde''} in the figure).
That the total $m_{1}$ moment satisfies a model independent sum rule
is clearly evident in the figure (the minor sensitivity is due to
differences in the TRK enhancement factors).  Instead, the total
dipole polarizability is unprotected by a sum rule and changes by
about 25\% over the range of values span by $R_{n}\!-\!R_{p}$. This
sensitivity and the ensuing strong correlation that emerges between
$\alpha_{D}$ and $R_{n}\!-\!R_{p}$ is consistent with the results
reported in Refs.~\cite{Satula:2005hy,Reinhard:2010wz}.  However, at
least for the class of models employed in this work, the sensitivity
of the Pygmy resonance to the neutron-skin thickness of ${}^{208}$Pb
is even higher---nearly 50\%---for both of the moments.

\begin{figure}[h]
\vspace{-0.05in}
\includegraphics[width=0.49\columnwidth,angle=0]{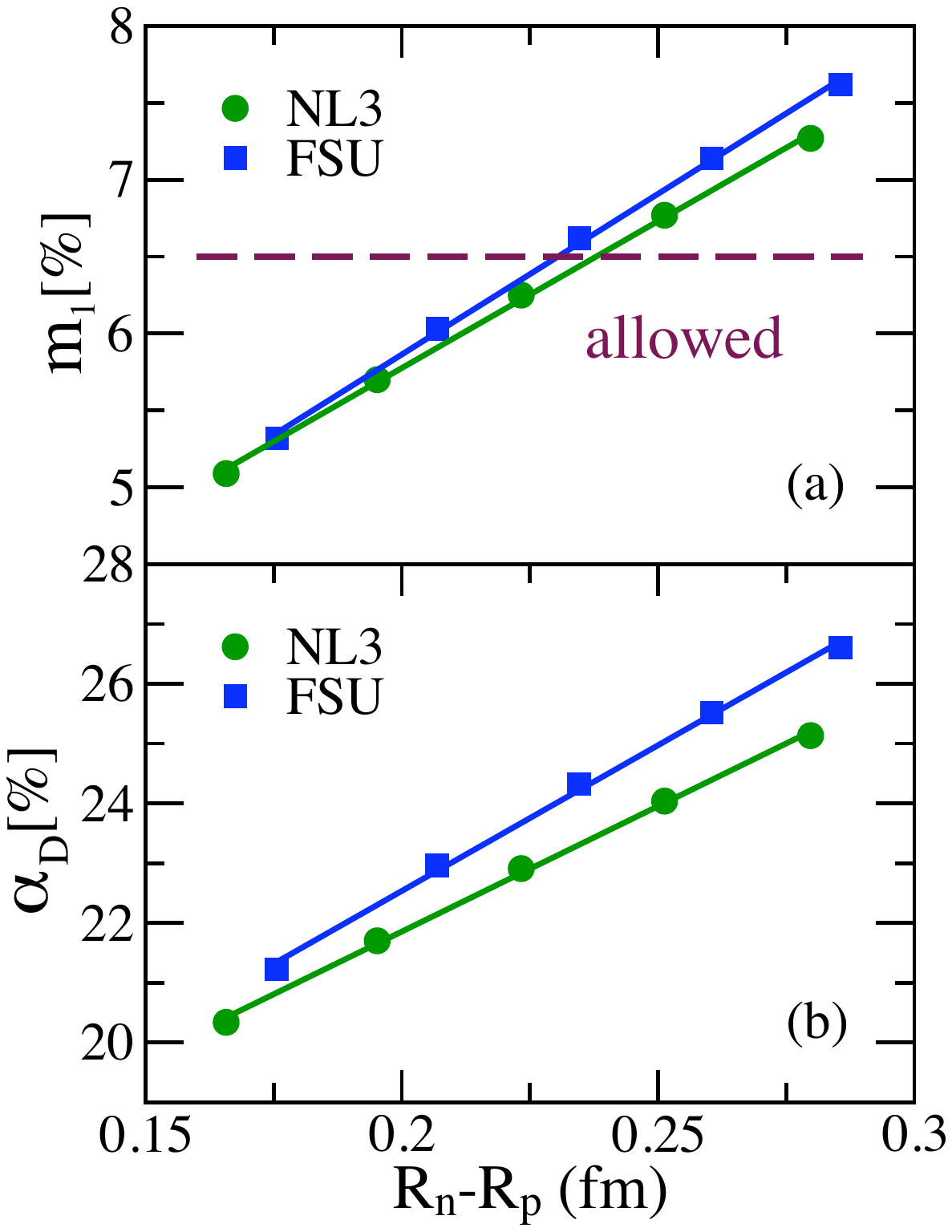}
\caption{(Color online) Percentage of the energy weighted sum rule 
 (a) and dipole polarizability (b) exhausted by the Pygmy dipole
 resonance in ${}^{68}$Ni as a function of the neutron-skin thickness 
  of ${}^{208}$Pb.}
\label{Fig7}
\end{figure}

We closed this section by displaying in Fig.~\ref{Fig7} the percentage
of the EWSR and of the dipole polarizability exhausted by the Pygmy
resonance in ${}^{68}$Ni. In both cases we find these quantities to be
strongly correlated to the neutron-skin thickness of ${}^{208}$Pb.
The dashed line in Fig.~\ref{Fig7}(a) represents an upper limit on the
fraction of the EWSR of about 6.5\% extracted from the analysis of
Carbone and collaborators~\cite{Carbone:2010az}. When combined with
the corresponding measurement on
${}^{132}$Sn~\cite{Adrich:2005,Klimkiewicz:2007zz}, the same analysis
reports values for the neutron skin thickness of ${}^{208}$Pb,
${}^{68}$Ni, and ${}^{132}$Sn of
$R_{n}\!-\!R_{p}\!=\!0.194\!\pm\!0.024$~fm, $0.200\!\pm\!0.015$~fm,
and $0.258\!\pm\!0.024$~fm, respectively~\cite{Carbone:2010az}.  Note
that the accurately-calibrated FSUGold parameter set predicts
neutron-skin thickness for these three nuclei of 0.207~fm, 0.211~fm,
and 0.271~fm, respectively, which fit comfortably within the above
limits.  However, although strongly correlated, the FSUGold
predictions appear consistently higher than the central values
suggested above~\cite{Carbone:2010az}. This may suggest a symmetry
energy even slightly softer than the one predicted by
FSUGold. Interestingly enough, astrophysical constraints emerging from
the study of neutron-star radii seem to support such a mild
softening~\cite{Steiner:2010fz,Fattoyev:2010mx}.

\begin{center}
\begin{table}
\begin{tabular}{|l||c|c|c|c|c|}
  \hline
   Model & $\Lambda_{\rm v}$ & $E_{c}$~(MeV) &
   $m_{1}$~(fm${}^{2}$MeV) & $m_{0}$~(fm${}^{2}$) & 
   $\alpha_{D}$~(fm${}^{3}$) \\ 
  \hline
  \hline
   NL3    
   & 0.00 & 17.563 & 263.868~(92.732) & 15.024~(86.206) & 3.531~(74.867) \\
   & 0.01 & 17.752 & 264.959~(93.233) & 14.926~(87.027) & 3.471~(75.960) \\
   & 0.02 & 17.958 & 265.608~(93.748) & 14.790~(87.870) & 3.402~(77.093) \\
   & 0.03 & 18.203 & 265.875~(94.299) & 14.606~(88.771) & 3.316~(78.302) \\
   & 0.04 & 18.503 & 265.502~(94.907) & 14.349~(89.773) & 3.207~(79.656) \\
  \hline
   FSU
   & 0.00 & 17.763 & 261.772~(92.380) & 14.737~(85.334) & 3.423~(73.399) \\
   & 0.01 & 17.955 & 262.482~(92.859) & 14.619~(86.117) & 3.360~(74.479) \\
   & 0.02 & 18.179 & 262.907~(93.383) & 14.462~(86.977) & 3.284~(75.678) \\
   & 0.03 & 18.454 & 262.914~(93.970) & 14.247~(87.946) & 3.188~(77.043) \\
   & 0.04 & 18.808 & 262.125~(94.679) & 13.937~(89.139) & 3.062~(78.766) \\
\hline
\end{tabular}
\caption{Contribution from the Giant-resonance region 
 ($11.25\!\le\!\omega\!\le 30$~MeV) to the various moments of 
 the distribution of dipole strength. The centroid energy 
 has been defined as $E_{c}\!=\!m_{1}/m_{0}$ and the quantities in
 parenthesis denote the fraction of the total moment contained in 
 the region of the Giant resonance.}
\label{Table5}
\end{table}
\end{center}

\section{Conclusions}
\label{sec:conclusions}

Motivated by two recent publications---one
experimental~\cite{Wieland:2009} and one
theoretical~\cite{Reinhard:2010wz}---the distribution of electric 
dipole strength in the neutron-rich ${}^{68}$Ni isotope was computed
using a relativistic RPA approach.  Concerning the experimental work,
Pygmy dipole strength carrying about 5\% of the energy weighted sum
rule was identified below the main GDR peak~\cite{Wieland:2009}.  This
result is significant as it complements earlier work on the
neutron-rich Tin isotopes that suggests a correlation between the
fraction of the TRK sum rule exhausted by the Pygmy resonance and the
neutron-skin thickness of
${}^{208}$Pb~\cite{Adrich:2005,Piekarewicz:2006ip}. Thus, one of the
goals of present work was to use the ${}^{68}$Ni measurement to
validate our earlier claim that models with overly large neutron skins
are in conflict with experiment~\cite{Piekarewicz:2006ip}.  In regard
to the recent theoretical work by Reinhard and Nazarewicz, we find it
interesting as it has established a strong correlation between the
dipole polarizability and the neutron-skin thickness of
${}^{208}$Pb~\cite{Reinhard:2010wz}.  As interesting but more
intriguing is the claim that the neutron-skin thickness of
${}^{208}$Pb is very weakly correlated with the low-energy electric
dipole strength. Such a claim is particularly intriguing given that
the dipole polarizability is proportional to the {\sl inverse energy
weighted sum} and, as such, the Pygmy resonance should exhaust a large
fraction of it.  Thus, explaining how can the neutron-skin thickness
of ${}^{208}$Pb be strongly correlated to the dipole polarizability
but weakly correlated to the Pygmy resonance became the second major
quest of this project.

To address these two issues we relied on a variety of effective
interactions that span a wide range of values for the neutron-skin
thickness of ${}^{208}$Pb. These effective interactions were derived
from accurately calibrated models that were suitably modified by
following a procedure first outlined in Ref.~\cite{Horowitz:2000xj}.
Such a procedure enables one to modify the density dependence of the
symmetry energy without compromising the success of the models in
describing well constrained nuclear observables.

We have used these models to generate the distribution of dipole
strength in ${}^{68}$Ni as well as various moments of the
distribution.  In particular, values in the 5-8\% range were generated
for the fraction of the energy weighted sum rule exhausted by the
Pygmy dipole resonance.  These values seem to fit comfortably within
the experimental range reported in Ref.~\cite{Wieland:2009}. Hence, by
itself, the measurement of low-energy dipole strength in ${}^{68}$Ni
does not seem to impose a stringent constraint on the density
dependence of the symmetry energy. However, when combined with an
earlier experiment on the neutron-rich Tin
isotopes~\cite{Adrich:2005}, the theoretical analysis presented in
Ref.~\cite{Carbone:2010az} argues for a tight constrain on the slope
of the symmetry energy ($L$) that excludes models with both very stiff
and very soft symmetry energies. It is significant that such an
analysis appears consistent with other approaches---based on
nuclear-structure and heavy-ion experiments---that have also been used
to constrain $L$. We note that to these approaches one can add
constrains obtained from both low- and high-density physics. Indeed,
the equation of state of dilute neutron matter (see
Refs.~\cite{Gezerlis:2009iw,Piekarewicz:2009gb} and references
therein) as well as neutron-star
radii~\cite{Steiner:2010fz,Fattoyev:2010mx} also favor values of $L$
within the range reported in Ref.~\cite{Carbone:2010az}.  In summary,
whereas the measurement of the PDR in ${}^{68}$Ni by itself does not
impose any stringent constrain on the density dependence of the
symmetry energy, it adds consistency to a picture that supports our
earlier claim that models with overly large neutron skins are in
conflict with experiment~\cite{Piekarewicz:2006ip}.

The dipole polarizability played center stage in this contribution
because of the expectation that it may act as a surrogate for the
neutron skin.  Indeed, semi-classical arguments have been used to
establish a direct correlation between the neutron-skin thickness and
the dipole polarizability~\cite{Satula:2005hy}. From our perspective,
this correlation emerges from the realization that the symmetry energy
acts as the restoring force of isovector oscillations. As such, models
with a soft symmetry energy---and thus a strong restoring
force---generate distributions of dipole strength that are both
hardened and quenched relative to their stiffer counterparts. And
whereas these two effects largely cancel out in the case of the energy
weighted sum (as they must in order to satisfy the sum rule) they work
for each other in the case of the dipole polarizability. This produces
a strong linear correlation between the dipole polarizability of
${}^{68}$Ni and the neutron-skin thickness of ${}^{208}$Pb---thereby
confirming the assertion of Ref.~\cite{Reinhard:2010wz}. But do we
also support the claim of a weak correlation between the neutron-skin
thickness of ${}^{208}$Pb and low-energy dipole strength? Quite the
contrary. To the extent that we can focus on the dipole
polarizability, we found a correlation just as strong and an even
larger sensitivity. This appears to be a natural consequence of the
following two facts: (a) the Pygmy resonance accounts for about
20-25\% of the dipole polarizability and (b) the neutron-skin
thickness of ${}^{208}$Pb is strongly correlated---with a correlation
coefficient of nearly one---to the total dipole polarizability of
${}^{68}$Ni~\cite{Reinhard:2010wz}. Note that in our earlier work on
the Tin isotopes the centroid energy of the PDR was found to be
insensitive to the density dependence of the symmetry
energy~\cite{Piekarewicz:2006ip}. There we found, as we do now,
centroid energies that are within 2\% of each other.  This should
be contrasted to the nearly 50\% sensitivity displayed by the fraction
of both the dipole polarizability and EWSR exhausted by the Pygmy
resonance. So in an attempt to confirm the assertions of
Ref.~\cite{Reinhard:2010wz} we obtained mixed results. On the one
hand, we confirmed the strong correlation between the dipole
polarizability in ${}^{68}$Ni and the neutron-skin thickness of
${}^{208}$Pb. On the other hand, we challenge the view---at least in
regard to the dipole polarizability---that the neutron-skin thickness
of ${}^{208}$Pb is very weakly correlated to the low-energy dipole
strength. Indeed, we suggest that the electromagnetic excitation of
both the Pygmy and Giant dipole resonances will continue to provide
powerful constraints on the density dependence of the symmetry energy.

\begin{acknowledgments}
 This work was supported in part by DOE grant DE-FG05-92ER40750.
\end{acknowledgments}

\bibliography{../ReferencesJP}

\begin{thebibliography}{63}
\expandafter\ifx\csname natexlab\endcsname\relax\def\natexlab#1{#1}\fi
\expandafter\ifx\csname bibnamefont\endcsname\relax
  \def\bibnamefont#1{#1}\fi
\expandafter\ifx\csname bibfnamefont\endcsname\relax
  \def\bibfnamefont#1{#1}\fi
\expandafter\ifx\csname citenamefont\endcsname\relax
  \def\citenamefont#1{#1}\fi
\expandafter\ifx\csname url\endcsname\relax
  \def\url#1{\texttt{#1}}\fi
\expandafter\ifx\csname urlprefix\endcsname\relax\def\urlprefix{URL }\fi
\providecommand{\bibinfo}[2]{#2}
\providecommand{\eprint}[2][]{\url{#2}}

\bibitem[{\citenamefont{Pollock et~al.}(1992)\citenamefont{Pollock, Fortson,
  and Wilets}}]{Pollock:1992mv}
\bibinfo{author}{\bibfnamefont{S.~J.} \bibnamefont{Pollock}},
  \bibinfo{author}{\bibfnamefont{E.~N.} \bibnamefont{Fortson}},
  \bibnamefont{and} \bibinfo{author}{\bibfnamefont{L.}~\bibnamefont{Wilets}},
  \bibinfo{journal}{Phys. Rev.} \textbf{\bibinfo{volume}{C46}},
  \bibinfo{pages}{2587} (\bibinfo{year}{1992}), \eprint{nucl-th/9211004}.

\bibitem[{\citenamefont{Sil et~al.}(2005)\citenamefont{Sil, Centelles, Vinas,
  and Piekarewicz}}]{Sil:2005tg}
\bibinfo{author}{\bibfnamefont{T.}~\bibnamefont{Sil}},
  \bibinfo{author}{\bibfnamefont{M.}~\bibnamefont{Centelles}},
  \bibinfo{author}{\bibfnamefont{X.}~\bibnamefont{Vinas}}, \bibnamefont{and}
  \bibinfo{author}{\bibfnamefont{J.}~\bibnamefont{Piekarewicz}},
  \bibinfo{journal}{Phys. Rev.} \textbf{\bibinfo{volume}{C71}},
  \bibinfo{pages}{045502} (\bibinfo{year}{2005}), \eprint{nucl-th/0501014}.

\bibitem[{\citenamefont{Brown}(2000)}]{Brown:2000}
\bibinfo{author}{\bibfnamefont{B.~A.} \bibnamefont{Brown}},
  \bibinfo{journal}{Phys. Rev. Lett.} \textbf{\bibinfo{volume}{85}},
  \bibinfo{pages}{5296} (\bibinfo{year}{2000}).

\bibitem[{\citenamefont{Furnstahl}(2002)}]{Furnstahl:2001un}
\bibinfo{author}{\bibfnamefont{R.~J.} \bibnamefont{Furnstahl}},
  \bibinfo{journal}{Nucl. Phys.} \textbf{\bibinfo{volume}{A706}},
  \bibinfo{pages}{85} (\bibinfo{year}{2002}), \eprint{nucl-th/0112085}.

\bibitem[{\citenamefont{Danielewicz}(2003)}]{Danielewicz:2003dd}
\bibinfo{author}{\bibfnamefont{P.}~\bibnamefont{Danielewicz}},
  \bibinfo{journal}{Nucl.Phys.} \textbf{\bibinfo{volume}{A727}},
  \bibinfo{pages}{233} (\bibinfo{year}{2003}), \eprint{nucl-th/0301050}.

\bibitem[{\citenamefont{Centelles et~al.}(2009)\citenamefont{Centelles,
  Roca-Maza, Vinas, and Warda}}]{Centelles:2008vu}
\bibinfo{author}{\bibfnamefont{M.}~\bibnamefont{Centelles}},
  \bibinfo{author}{\bibfnamefont{X.}~\bibnamefont{Roca-Maza}},
  \bibinfo{author}{\bibfnamefont{X.}~\bibnamefont{Vinas}}, \bibnamefont{and}
  \bibinfo{author}{\bibfnamefont{M.}~\bibnamefont{Warda}},
  \bibinfo{journal}{Phys.Rev.Lett.} \textbf{\bibinfo{volume}{102}},
  \bibinfo{pages}{122502} (\bibinfo{year}{2009}), \eprint{0806.2886}.

\bibitem[{\citenamefont{Centelles et~al.}(2010)\citenamefont{Centelles,
  Roca-Maza, Vinas, and Warda}}]{Centelles:2010qh}
\bibinfo{author}{\bibfnamefont{M.}~\bibnamefont{Centelles}},
  \bibinfo{author}{\bibfnamefont{X.}~\bibnamefont{Roca-Maza}},
  \bibinfo{author}{\bibfnamefont{X.}~\bibnamefont{Vinas}}, \bibnamefont{and}
  \bibinfo{author}{\bibfnamefont{M.}~\bibnamefont{Warda}},
  \bibinfo{journal}{Phys.Rev.} \textbf{\bibinfo{volume}{C82}},
  \bibinfo{pages}{054314} (\bibinfo{year}{2010}), \eprint{1010.5396}.

\bibitem[{\citenamefont{Tsang et~al.}(2004)}]{Tsang:2004zz}
\bibinfo{author}{\bibfnamefont{M.~B.} \bibnamefont{Tsang}}
  \bibnamefont{et~al.}, \bibinfo{journal}{Phys. Rev. Lett.}
  \textbf{\bibinfo{volume}{92}}, \bibinfo{pages}{062701}
  (\bibinfo{year}{2004}).

\bibitem[{\citenamefont{Chen et~al.}(2005)\citenamefont{Chen, Ko, and
  Li}}]{Chen:2004si}
\bibinfo{author}{\bibfnamefont{L.-W.} \bibnamefont{Chen}},
  \bibinfo{author}{\bibfnamefont{C.~M.} \bibnamefont{Ko}}, \bibnamefont{and}
  \bibinfo{author}{\bibfnamefont{B.-A.} \bibnamefont{Li}},
  \bibinfo{journal}{Phys. Rev. Lett.} \textbf{\bibinfo{volume}{94}},
  \bibinfo{pages}{032701} (\bibinfo{year}{2005}), \eprint{nucl-th/0407032}.

\bibitem[{\citenamefont{Steiner and Li}(2005)}]{Steiner:2005rd}
\bibinfo{author}{\bibfnamefont{A.~W.} \bibnamefont{Steiner}} \bibnamefont{and}
  \bibinfo{author}{\bibfnamefont{B.-A.} \bibnamefont{Li}},
  \bibinfo{journal}{Phys. Rev.} \textbf{\bibinfo{volume}{C72}},
  \bibinfo{pages}{041601} (\bibinfo{year}{2005}), \eprint{nucl-th/0505051}.

\bibitem[{\citenamefont{Shetty et~al.}(2007)\citenamefont{Shetty, Yennello, and
  Souliotis}}]{Shetty:2007zg}
\bibinfo{author}{\bibfnamefont{D.~V.} \bibnamefont{Shetty}},
  \bibinfo{author}{\bibfnamefont{S.~J.} \bibnamefont{Yennello}},
  \bibnamefont{and} \bibinfo{author}{\bibfnamefont{G.~A.}
  \bibnamefont{Souliotis}}, \bibinfo{journal}{Phys. Rev.}
  \textbf{\bibinfo{volume}{C76}}, \bibinfo{pages}{024606}
  (\bibinfo{year}{2007}), \eprint{0704.0471}.

\bibitem[{\citenamefont{Tsang et~al.}(2009)}]{Tsang:2008fd}
\bibinfo{author}{\bibfnamefont{M.~B.} \bibnamefont{Tsang}}
  \bibnamefont{et~al.}, \bibinfo{journal}{Phys. Rev. Lett.}
  \textbf{\bibinfo{volume}{102}}, \bibinfo{pages}{122701}
  (\bibinfo{year}{2009}), \eprint{0811.3107}.

\bibitem[{\citenamefont{Horowitz and
  Piekarewicz}(2001{\natexlab{a}})}]{Horowitz:2000xj}
\bibinfo{author}{\bibfnamefont{C.~J.} \bibnamefont{Horowitz}} \bibnamefont{and}
  \bibinfo{author}{\bibfnamefont{J.}~\bibnamefont{Piekarewicz}},
  \bibinfo{journal}{Phys. Rev. Lett.} \textbf{\bibinfo{volume}{86}},
  \bibinfo{pages}{5647} (\bibinfo{year}{2001}{\natexlab{a}}),
  \eprint{astro-ph/0010227}.

\bibitem[{\citenamefont{Horowitz and
  Piekarewicz}(2001{\natexlab{b}})}]{Horowitz:2001ya}
\bibinfo{author}{\bibfnamefont{C.~J.} \bibnamefont{Horowitz}} \bibnamefont{and}
  \bibinfo{author}{\bibfnamefont{J.}~\bibnamefont{Piekarewicz}},
  \bibinfo{journal}{Phys. Rev.} \textbf{\bibinfo{volume}{C64}},
  \bibinfo{pages}{062802} (\bibinfo{year}{2001}{\natexlab{b}}),
  \eprint{nucl-th/0108036}.

\bibitem[{\citenamefont{Horowitz and Piekarewicz}(2002)}]{Horowitz:2002mb}
\bibinfo{author}{\bibfnamefont{C.~J.} \bibnamefont{Horowitz}} \bibnamefont{and}
  \bibinfo{author}{\bibfnamefont{J.}~\bibnamefont{Piekarewicz}},
  \bibinfo{journal}{Phys. Rev.} \textbf{\bibinfo{volume}{C66}},
  \bibinfo{pages}{055803} (\bibinfo{year}{2002}), \eprint{nucl-th/0207067}.

\bibitem[{\citenamefont{Carriere et~al.}(2003)\citenamefont{Carriere, Horowitz,
  and Piekarewicz}}]{Carriere:2002bx}
\bibinfo{author}{\bibfnamefont{J.}~\bibnamefont{Carriere}},
  \bibinfo{author}{\bibfnamefont{C.~J.} \bibnamefont{Horowitz}},
  \bibnamefont{and}
  \bibinfo{author}{\bibfnamefont{J.}~\bibnamefont{Piekarewicz}},
  \bibinfo{journal}{Astrophys. J.} \textbf{\bibinfo{volume}{593}},
  \bibinfo{pages}{463} (\bibinfo{year}{2003}), \eprint{nucl-th/0211015}.

\bibitem[{\citenamefont{Steiner et~al.}(2005)\citenamefont{Steiner, Prakash,
  Lattimer, and Ellis}}]{Steiner:2004fi}
\bibinfo{author}{\bibfnamefont{A.~W.} \bibnamefont{Steiner}},
  \bibinfo{author}{\bibfnamefont{M.}~\bibnamefont{Prakash}},
  \bibinfo{author}{\bibfnamefont{J.~M.} \bibnamefont{Lattimer}},
  \bibnamefont{and} \bibinfo{author}{\bibfnamefont{P.~J.} \bibnamefont{Ellis}},
  \bibinfo{journal}{Phys. Rept.} \textbf{\bibinfo{volume}{411}},
  \bibinfo{pages}{325} (\bibinfo{year}{2005}), \eprint{nucl-th/0410066}.

\bibitem[{\citenamefont{Li and Steiner}(2006)}]{Li:2005sr}
\bibinfo{author}{\bibfnamefont{B.-A.} \bibnamefont{Li}} \bibnamefont{and}
  \bibinfo{author}{\bibfnamefont{A.~W.} \bibnamefont{Steiner}},
  \bibinfo{journal}{Phys. Lett.} \textbf{\bibinfo{volume}{B642}},
  \bibinfo{pages}{436} (\bibinfo{year}{2006}), \eprint{nucl-th/0511064}.

\bibitem[{\citenamefont{Ray and Hoffmann}(1985)}]{Ray:1985yg}
\bibinfo{author}{\bibfnamefont{L.}~\bibnamefont{Ray}} \bibnamefont{and}
  \bibinfo{author}{\bibfnamefont{G.~W.} \bibnamefont{Hoffmann}},
  \bibinfo{journal}{Phys. Rev.} \textbf{\bibinfo{volume}{C31}},
  \bibinfo{pages}{538} (\bibinfo{year}{1985}).

\bibitem[{\citenamefont{Ray et~al.}(1992)\citenamefont{Ray, Hoffmann, and
  Coker}}]{Ray:1992fj}
\bibinfo{author}{\bibfnamefont{L.}~\bibnamefont{Ray}},
  \bibinfo{author}{\bibfnamefont{G.~W.} \bibnamefont{Hoffmann}},
  \bibnamefont{and} \bibinfo{author}{\bibfnamefont{W.~R.} \bibnamefont{Coker}},
  \bibinfo{journal}{Phys. Rept.} \textbf{\bibinfo{volume}{212}},
  \bibinfo{pages}{223} (\bibinfo{year}{1992}).

\bibitem[{\citenamefont{Horowitz et~al.}(2001)\citenamefont{Horowitz, Pollock,
  Souder, and Michaels}}]{Horowitz:1999fk}
\bibinfo{author}{\bibfnamefont{C.~J.} \bibnamefont{Horowitz}},
  \bibinfo{author}{\bibfnamefont{S.~J.} \bibnamefont{Pollock}},
  \bibinfo{author}{\bibfnamefont{P.~A.} \bibnamefont{Souder}},
  \bibnamefont{and} \bibinfo{author}{\bibfnamefont{R.}~\bibnamefont{Michaels}},
  \bibinfo{journal}{Phys. Rev.} \textbf{\bibinfo{volume}{C63}},
  \bibinfo{pages}{025501} (\bibinfo{year}{2001}), \eprint{nucl-th/9912038}.

\bibitem[{\citenamefont{Kumar et~al.}(2005)\citenamefont{Kumar, Michaels,
  Souder, and Urciuoli}}]{Michaels:2005}
\bibinfo{author}{\bibfnamefont{K.}~\bibnamefont{Kumar}},
  \bibinfo{author}{\bibfnamefont{R.}~\bibnamefont{Michaels}},
  \bibinfo{author}{\bibfnamefont{P.~A.} \bibnamefont{Souder}},
  \bibnamefont{and} \bibinfo{author}{\bibfnamefont{G.~M.}
  \bibnamefont{Urciuoli}} (\bibinfo{year}{2005}),
  \urlprefix\url{http://hallaweb.jlab.org/parity/prex}.

\bibitem[{\citenamefont{Harakeh and van~der Woude}(2001)}]{Harakeh:2001}
\bibinfo{author}{\bibfnamefont{M.~N.} \bibnamefont{Harakeh}} \bibnamefont{and}
  \bibinfo{author}{\bibfnamefont{A.}~\bibnamefont{van~der Woude}},
  \emph{\bibinfo{title}{Giant Resonances-Fundamental High-frequency Modes of
  Nuclear Excitation}} (\bibinfo{publisher}{Clarendon, Oxford},
  \bibinfo{year}{2001}).

\bibitem[{\citenamefont{Suzuki et~al.}(1990)\citenamefont{Suzuki, Ikeda, and
  Sato}}]{Suzuki:1990}
\bibinfo{author}{\bibfnamefont{Y.}~\bibnamefont{Suzuki}},
  \bibinfo{author}{\bibfnamefont{K.}~\bibnamefont{Ikeda}}, \bibnamefont{and}
  \bibinfo{author}{\bibfnamefont{H.}~\bibnamefont{Sato}},
  \bibinfo{journal}{Prog. Theor. Phys.} \textbf{\bibinfo{volume}{83}},
  \bibinfo{pages}{180} (\bibinfo{year}{1990}).

\bibitem[{\citenamefont{Van~Isacker and Nagarajan}(1992)}]{VanIsacker:1992}
\bibinfo{author}{\bibfnamefont{P.}~\bibnamefont{Van~Isacker}} \bibnamefont{and}
  \bibinfo{author}{\bibfnamefont{D.~D.} \bibnamefont{Nagarajan},
  \bibfnamefont{M.~A.and~Warner}}, \bibinfo{journal}{Phys. Rev.}
  \textbf{\bibinfo{volume}{C45}}, \bibinfo{pages}{R13} (\bibinfo{year}{1992}).

\bibitem[{\citenamefont{Hamamoto et~al.}(1996)\citenamefont{Hamamoto, Sagawa,
  and Zhang}}]{Hamamoto:1996}
\bibinfo{author}{\bibfnamefont{I.}~\bibnamefont{Hamamoto}},
  \bibinfo{author}{\bibfnamefont{H.}~\bibnamefont{Sagawa}}, \bibnamefont{and}
  \bibinfo{author}{\bibfnamefont{X.~Z.} \bibnamefont{Zhang}},
  \bibinfo{journal}{Phys. Rev.} \textbf{\bibinfo{volume}{C53}},
  \bibinfo{pages}{765} (\bibinfo{year}{1996}).

\bibitem[{\citenamefont{Hamamoto et~al.}(1998)\citenamefont{Hamamoto, Sagawa,
  and Zhang}}]{Hamamoto:1998}
\bibinfo{author}{\bibfnamefont{I.}~\bibnamefont{Hamamoto}},
  \bibinfo{author}{\bibfnamefont{H.}~\bibnamefont{Sagawa}}, \bibnamefont{and}
  \bibinfo{author}{\bibfnamefont{X.~Z.} \bibnamefont{Zhang}},
  \bibinfo{journal}{Phys. Rev.} \textbf{\bibinfo{volume}{C57}},
  \bibinfo{pages}{R1064} (\bibinfo{year}{1998}).

\bibitem[{\citenamefont{Vretenar
  et~al.}(2001{\natexlab{a}})\citenamefont{Vretenar, Paar, Ring, and
  Lalazissis}}]{Vretenar:2000yy}
\bibinfo{author}{\bibfnamefont{D.}~\bibnamefont{Vretenar}},
  \bibinfo{author}{\bibfnamefont{N.}~\bibnamefont{Paar}},
  \bibinfo{author}{\bibfnamefont{P.}~\bibnamefont{Ring}}, \bibnamefont{and}
  \bibinfo{author}{\bibfnamefont{G.~A.} \bibnamefont{Lalazissis}},
  \bibinfo{journal}{Phys. Rev.} \textbf{\bibinfo{volume}{C63}},
  \bibinfo{pages}{047301} (\bibinfo{year}{2001}{\natexlab{a}}),
  \eprint{nucl-th/0009057}.

\bibitem[{\citenamefont{Vretenar
  et~al.}(2001{\natexlab{b}})\citenamefont{Vretenar, Paar, Ring, and
  Lalazissis}}]{Vretenar:2001hs}
\bibinfo{author}{\bibfnamefont{D.}~\bibnamefont{Vretenar}},
  \bibinfo{author}{\bibfnamefont{N.}~\bibnamefont{Paar}},
  \bibinfo{author}{\bibfnamefont{P.}~\bibnamefont{Ring}}, \bibnamefont{and}
  \bibinfo{author}{\bibfnamefont{G.~A.} \bibnamefont{Lalazissis}},
  \bibinfo{journal}{Nucl. Phys.} \textbf{\bibinfo{volume}{A692}},
  \bibinfo{pages}{496} (\bibinfo{year}{2001}{\natexlab{b}}),
  \eprint{nucl-th/0101063}.

\bibitem[{\citenamefont{Paar et~al.}(2005)\citenamefont{Paar, Niksic, Vretenar,
  and Ring}}]{Paar:2004gr}
\bibinfo{author}{\bibfnamefont{N.}~\bibnamefont{Paar}},
  \bibinfo{author}{\bibfnamefont{T.}~\bibnamefont{Niksic}},
  \bibinfo{author}{\bibfnamefont{D.}~\bibnamefont{Vretenar}}, \bibnamefont{and}
  \bibinfo{author}{\bibfnamefont{P.}~\bibnamefont{Ring}},
  \bibinfo{journal}{Phys. Lett.} \textbf{\bibinfo{volume}{B606}},
  \bibinfo{pages}{288} (\bibinfo{year}{2005}), \eprint{nucl-th/0404055}.

\bibitem[{\citenamefont{Piekarewicz}(2006)}]{Piekarewicz:2006ip}
\bibinfo{author}{\bibfnamefont{J.}~\bibnamefont{Piekarewicz}},
  \bibinfo{journal}{Phys. Rev.} \textbf{\bibinfo{volume}{C73}},
  \bibinfo{pages}{044325} (\bibinfo{year}{2006}), \eprint{nucl-th/0602036}.

\bibitem[{\citenamefont{Tsoneva et~al.}(2004)\citenamefont{Tsoneva, Lenske, and
  Stoyanov}}]{Tsoneva:2003gv}
\bibinfo{author}{\bibfnamefont{N.}~\bibnamefont{Tsoneva}},
  \bibinfo{author}{\bibfnamefont{H.}~\bibnamefont{Lenske}}, \bibnamefont{and}
  \bibinfo{author}{\bibfnamefont{C.}~\bibnamefont{Stoyanov}},
  \bibinfo{journal}{Phys. Lett.} \textbf{\bibinfo{volume}{B586}},
  \bibinfo{pages}{213} (\bibinfo{year}{2004}), \eprint{nucl-th/0307020}.

\bibitem[{\citenamefont{Tsoneva and Lenske}(2008)}]{Tsoneva:2007fk}
\bibinfo{author}{\bibfnamefont{N.}~\bibnamefont{Tsoneva}} \bibnamefont{and}
  \bibinfo{author}{\bibfnamefont{H.}~\bibnamefont{Lenske}},
  \bibinfo{journal}{Phys. Rev.} \textbf{\bibinfo{volume}{C77}},
  \bibinfo{pages}{024321} (\bibinfo{year}{2008}), \eprint{0706.4204}.

\bibitem[{\citenamefont{Klimkiewicz et~al.}(2007)}]{Klimkiewicz:2007zz}
\bibinfo{author}{\bibfnamefont{A.}~\bibnamefont{Klimkiewicz}}
  \bibnamefont{et~al.}, \bibinfo{journal}{Phys. Rev.}
  \textbf{\bibinfo{volume}{C76}}, \bibinfo{pages}{051603}
  (\bibinfo{year}{2007}).

\bibitem[{\citenamefont{Carbone et~al.}(2010)\citenamefont{Carbone, Colo,
  Bracco, Cao, Bortignon et~al.}}]{Carbone:2010az}
\bibinfo{author}{\bibfnamefont{A.}~\bibnamefont{Carbone}},
  \bibinfo{author}{\bibfnamefont{G.}~\bibnamefont{Colo}},
  \bibinfo{author}{\bibfnamefont{A.}~\bibnamefont{Bracco}},
  \bibinfo{author}{\bibfnamefont{L.-G.} \bibnamefont{Cao}},
  \bibinfo{author}{\bibfnamefont{P.~F.} \bibnamefont{Bortignon}},
  \bibnamefont{et~al.}, \bibinfo{journal}{Phys.Rev.}
  \textbf{\bibinfo{volume}{C81}}, \bibinfo{pages}{041301}
  (\bibinfo{year}{2010}), \eprint{1003.3580}.

\bibitem[{\citenamefont{Adrich et~al.}(2005)}]{Adrich:2005}
\bibinfo{author}{\bibfnamefont{P.}~\bibnamefont{Adrich}} \bibnamefont{et~al.},
  \bibinfo{journal}{Phys. Rev. Lett} \textbf{\bibinfo{volume}{95}},
  \bibinfo{pages}{132501} (\bibinfo{year}{2005}).

\bibitem[{\citenamefont{Wieland et~al.}(2009)}]{Wieland:2009}
\bibinfo{author}{\bibfnamefont{O.}~\bibnamefont{Wieland}} \bibnamefont{et~al.},
  \bibinfo{journal}{Phys. Rev. Lett.} \textbf{\bibinfo{volume}{102}},
  \bibinfo{pages}{092502} (\bibinfo{year}{2009}).

\bibitem[{\citenamefont{Reinhard and Nazarewicz}(2010)}]{Reinhard:2010wz}
\bibinfo{author}{\bibfnamefont{P.-G.} \bibnamefont{Reinhard}} \bibnamefont{and}
  \bibinfo{author}{\bibfnamefont{W.}~\bibnamefont{Nazarewicz}},
  \bibinfo{journal}{Phys.Rev.} \textbf{\bibinfo{volume}{C81}},
  \bibinfo{pages}{051303} (\bibinfo{year}{2010}), \eprint{1002.4140}.

\bibitem[{\citenamefont{Paar et~al.}(2007)\citenamefont{Paar, Vretenar, Khan,
  and Colo}}]{Paar:2007bk}
\bibinfo{author}{\bibfnamefont{N.}~\bibnamefont{Paar}},
  \bibinfo{author}{\bibfnamefont{D.}~\bibnamefont{Vretenar}},
  \bibinfo{author}{\bibfnamefont{E.}~\bibnamefont{Khan}}, \bibnamefont{and}
  \bibinfo{author}{\bibfnamefont{G.}~\bibnamefont{Colo}},
  \bibinfo{journal}{Rept. Prog. Phys.} \textbf{\bibinfo{volume}{70}},
  \bibinfo{pages}{691} (\bibinfo{year}{2007}), \eprint{nucl-th/0701081}.

\bibitem[{\citenamefont{Paar}(2010)}]{Paar:2010ww}
\bibinfo{author}{\bibfnamefont{N.}~\bibnamefont{Paar}}, \bibinfo{journal}{J.
  Phys.} \textbf{\bibinfo{volume}{G37}}, \bibinfo{pages}{064014}
  (\bibinfo{year}{2010}), \eprint{1002.4776}.

\bibitem[{\citenamefont{Mueller and Serot}(1996)}]{Mueller:1996pm}
\bibinfo{author}{\bibfnamefont{H.}~\bibnamefont{Mueller}} \bibnamefont{and}
  \bibinfo{author}{\bibfnamefont{B.~D.} \bibnamefont{Serot}},
  \bibinfo{journal}{Nucl. Phys.} \textbf{\bibinfo{volume}{A606}},
  \bibinfo{pages}{508} (\bibinfo{year}{1996}), \eprint{nucl-th/9603037}.

\bibitem[{\citenamefont{Serot and Walecka}(1986)}]{Serot:1984ey}
\bibinfo{author}{\bibfnamefont{B.~D.} \bibnamefont{Serot}} \bibnamefont{and}
  \bibinfo{author}{\bibfnamefont{J.~D.} \bibnamefont{Walecka}},
  \bibinfo{journal}{Adv. Nucl. Phys.} \textbf{\bibinfo{volume}{16}},
  \bibinfo{pages}{1} (\bibinfo{year}{1986}).

\bibitem[{\citenamefont{Serot and Walecka}(1997)}]{Serot:1997xg}
\bibinfo{author}{\bibfnamefont{B.~D.} \bibnamefont{Serot}} \bibnamefont{and}
  \bibinfo{author}{\bibfnamefont{J.~D.} \bibnamefont{Walecka}},
  \bibinfo{journal}{Int. J. Mod. Phys.} \textbf{\bibinfo{volume}{E6}},
  \bibinfo{pages}{515} (\bibinfo{year}{1997}), \eprint{nucl-th/9701058}.

\bibitem[{\citenamefont{Boguta and Bodmer}(1977)}]{Boguta:1977xi}
\bibinfo{author}{\bibfnamefont{J.}~\bibnamefont{Boguta}} \bibnamefont{and}
  \bibinfo{author}{\bibfnamefont{A.~R.} \bibnamefont{Bodmer}},
  \bibinfo{journal}{Nucl. Phys.} \textbf{\bibinfo{volume}{A292}},
  \bibinfo{pages}{413} (\bibinfo{year}{1977}).

\bibitem[{\citenamefont{Walecka}(1974)}]{Walecka:1974qa}
\bibinfo{author}{\bibfnamefont{J.~D.} \bibnamefont{Walecka}},
  \bibinfo{journal}{Annals Phys} \textbf{\bibinfo{volume}{83}},
  \bibinfo{pages}{491} (\bibinfo{year}{1974}).

\bibitem[{\citenamefont{Youngblood et~al.}(1999)\citenamefont{Youngblood,
  Clark, and Lui}}]{Youngblood:1999}
\bibinfo{author}{\bibfnamefont{D.~H.} \bibnamefont{Youngblood}},
  \bibinfo{author}{\bibfnamefont{H.~L.} \bibnamefont{Clark}}, \bibnamefont{and}
  \bibinfo{author}{\bibfnamefont{Y.-W.} \bibnamefont{Lui}},
  \bibinfo{journal}{Phys. Rev. Lett.} \textbf{\bibinfo{volume}{82}},
  \bibinfo{pages}{691} (\bibinfo{year}{1999}).

\bibitem[{\citenamefont{Piekarewicz}(2001)}]{Piekarewicz:2001nm}
\bibinfo{author}{\bibfnamefont{J.}~\bibnamefont{Piekarewicz}},
  \bibinfo{journal}{Phys. Rev.} \textbf{\bibinfo{volume}{C64}},
  \bibinfo{pages}{024307} (\bibinfo{year}{2001}), \eprint{nucl-th/0103016}.

\bibitem[{\citenamefont{Piekarewicz}(2002)}]{Piekarewicz:2002jd}
\bibinfo{author}{\bibfnamefont{J.}~\bibnamefont{Piekarewicz}},
  \bibinfo{journal}{Phys. Rev.} \textbf{\bibinfo{volume}{C66}},
  \bibinfo{pages}{034305} (\bibinfo{year}{2002}).

\bibitem[{\citenamefont{Col\`o et~al.}(2004)\citenamefont{Col\`o, Van~Giai,
  Meyer, Bennaceur, and Bonche}}]{Colo:2004mj}
\bibinfo{author}{\bibfnamefont{G.}~\bibnamefont{Col\`o}},
  \bibinfo{author}{\bibfnamefont{N.}~\bibnamefont{Van~Giai}},
  \bibinfo{author}{\bibfnamefont{J.}~\bibnamefont{Meyer}},
  \bibinfo{author}{\bibfnamefont{K.}~\bibnamefont{Bennaceur}},
  \bibnamefont{and} \bibinfo{author}{\bibfnamefont{P.}~\bibnamefont{Bonche}},
  \bibinfo{journal}{Phys. Rev.} \textbf{\bibinfo{volume}{C70}},
  \bibinfo{pages}{024307} (\bibinfo{year}{2004}), \eprint{nucl-th/0403086}.

\bibitem[{\citenamefont{Todd and Piekarewicz}(2003)}]{Todd:2003xs}
\bibinfo{author}{\bibfnamefont{B.~G.} \bibnamefont{Todd}} \bibnamefont{and}
  \bibinfo{author}{\bibfnamefont{J.}~\bibnamefont{Piekarewicz}},
  \bibinfo{journal}{Phys. Rev.} \textbf{\bibinfo{volume}{C67}},
  \bibinfo{pages}{044317} (\bibinfo{year}{2003}), \eprint{nucl-th/0301092}.

\bibitem[{\citenamefont{Fetter and Walecka}(1971)}]{Fetter:1971}
\bibinfo{author}{\bibfnamefont{A.~L.} \bibnamefont{Fetter}} \bibnamefont{and}
  \bibinfo{author}{\bibfnamefont{J.~D.} \bibnamefont{Walecka}},
  \emph{\bibinfo{title}{Quantum Theory of Many Particle Systems}}
  (\bibinfo{publisher}{McGraw-Hill, New York}, \bibinfo{year}{1971}).

\bibitem[{\citenamefont{Dawson and Furnstahl}(1990)}]{Dawson:1990wp}
\bibinfo{author}{\bibfnamefont{J.~F.} \bibnamefont{Dawson}} \bibnamefont{and}
  \bibinfo{author}{\bibfnamefont{R.~J.} \bibnamefont{Furnstahl}},
  \bibinfo{journal}{Phys. Rev.} \textbf{\bibinfo{volume}{C42}},
  \bibinfo{pages}{2009} (\bibinfo{year}{1990}).

\bibitem[{\citenamefont{Todd-Rutel and Piekarewicz}(2005)}]{Todd-Rutel:2005fa}
\bibinfo{author}{\bibfnamefont{B.~G.} \bibnamefont{Todd-Rutel}}
  \bibnamefont{and}
  \bibinfo{author}{\bibfnamefont{J.}~\bibnamefont{Piekarewicz}},
  \bibinfo{journal}{Phys. Rev. Lett} \textbf{\bibinfo{volume}{95}},
  \bibinfo{pages}{122501} (\bibinfo{year}{2005}), \eprint{nucl-th/0504034}.

\bibitem[{\citenamefont{Lalazissis et~al.}(1997)\citenamefont{Lalazissis,
  Konig, and Ring}}]{Lalazissis:1996rd}
\bibinfo{author}{\bibfnamefont{G.~A.} \bibnamefont{Lalazissis}},
  \bibinfo{author}{\bibfnamefont{J.}~\bibnamefont{Konig}}, \bibnamefont{and}
  \bibinfo{author}{\bibfnamefont{P.}~\bibnamefont{Ring}},
  \bibinfo{journal}{Phys. Rev.} \textbf{\bibinfo{volume}{C55}},
  \bibinfo{pages}{540} (\bibinfo{year}{1997}), \eprint{nucl-th/9607039}.

\bibitem[{\citenamefont{Lalazissis et~al.}(1999)\citenamefont{Lalazissis,
  Raman, and Ring}}]{Lalazissis:1999}
\bibinfo{author}{\bibfnamefont{G.~A.} \bibnamefont{Lalazissis}},
  \bibinfo{author}{\bibfnamefont{S.}~\bibnamefont{Raman}}, \bibnamefont{and}
  \bibinfo{author}{\bibfnamefont{P.}~\bibnamefont{Ring}}, \bibinfo{journal}{At.
  Data Nucl. Data Tables} \textbf{\bibinfo{volume}{71}}, \bibinfo{pages}{1}
  (\bibinfo{year}{1999}).

\bibitem[{\citenamefont{Danielewicz et~al.}(2002)\citenamefont{Danielewicz,
  Lacey, and Lynch}}]{Danielewicz:2002pu}
\bibinfo{author}{\bibfnamefont{P.}~\bibnamefont{Danielewicz}},
  \bibinfo{author}{\bibfnamefont{R.}~\bibnamefont{Lacey}}, \bibnamefont{and}
  \bibinfo{author}{\bibfnamefont{W.~G.} \bibnamefont{Lynch}},
  \bibinfo{journal}{Science} \textbf{\bibinfo{volume}{298}},
  \bibinfo{pages}{1592} (\bibinfo{year}{2002}), \eprint{nucl-th/0208016}.

\bibitem[{\citenamefont{Piekarewicz}(2004)}]{Piekarewicz:2003br}
\bibinfo{author}{\bibfnamefont{J.}~\bibnamefont{Piekarewicz}},
  \bibinfo{journal}{Phys. Rev.} \textbf{\bibinfo{volume}{C69}},
  \bibinfo{pages}{041301} (\bibinfo{year}{2004}), \eprint{nucl-th/0312020}.

\bibitem[{\citenamefont{Piekarewicz}(2010)}]{Piekarewicz:2009gb}
\bibinfo{author}{\bibfnamefont{J.}~\bibnamefont{Piekarewicz}},
  \bibinfo{journal}{J. Phys.} \textbf{\bibinfo{volume}{G37}},
  \bibinfo{pages}{064038} (\bibinfo{year}{2010}), \eprint{0912.5103}.

\bibitem[{\citenamefont{De~Vries et~al.}(1987)\citenamefont{De~Vries, De~Jager,
  and De~Vries}}]{DeJager:1987qc}
\bibinfo{author}{\bibfnamefont{H.}~\bibnamefont{De~Vries}},
  \bibinfo{author}{\bibfnamefont{C.~W.} \bibnamefont{De~Jager}},
  \bibnamefont{and} \bibinfo{author}{\bibfnamefont{C.}~\bibnamefont{De~Vries}},
  \bibinfo{journal}{Atom. Data Nucl. Data Tabl.} \textbf{\bibinfo{volume}{36}},
  \bibinfo{pages}{495} (\bibinfo{year}{1987}).

\bibitem[{\citenamefont{Satula et~al.}(2006)\citenamefont{Satula, Wyss, and
  Rafalski}}]{Satula:2005hy}
\bibinfo{author}{\bibfnamefont{W.}~\bibnamefont{Satula}},
  \bibinfo{author}{\bibfnamefont{R.~A.} \bibnamefont{Wyss}}, \bibnamefont{and}
  \bibinfo{author}{\bibfnamefont{M.}~\bibnamefont{Rafalski}},
  \bibinfo{journal}{Phys.Rev.} \textbf{\bibinfo{volume}{C74}},
  \bibinfo{pages}{011301} (\bibinfo{year}{2006}), \eprint{nucl-th/0508004}.

\bibitem[{\citenamefont{Steiner et~al.}(2010)\citenamefont{Steiner, Lattimer,
  and Brown}}]{Steiner:2010fz}
\bibinfo{author}{\bibfnamefont{A.~W.} \bibnamefont{Steiner}},
  \bibinfo{author}{\bibfnamefont{J.~M.} \bibnamefont{Lattimer}},
  \bibnamefont{and} \bibinfo{author}{\bibfnamefont{E.~F.} \bibnamefont{Brown}},
  \bibinfo{journal}{Astrophys.J.} \textbf{\bibinfo{volume}{722}},
  \bibinfo{pages}{33} (\bibinfo{year}{2010}), \bibinfo{note}{* Temporary entry
  *}, \eprint{1005.0811}.

\bibitem[{\citenamefont{Fattoyev et~al.}(2010)\citenamefont{Fattoyev, Horowitz,
  Piekarewicz, and Shen}}]{Fattoyev:2010mx}
\bibinfo{author}{\bibfnamefont{F.}~\bibnamefont{Fattoyev}},
  \bibinfo{author}{\bibfnamefont{C.}~\bibnamefont{Horowitz}},
  \bibinfo{author}{\bibfnamefont{J.}~\bibnamefont{Piekarewicz}},
  \bibnamefont{and} \bibinfo{author}{\bibfnamefont{G.}~\bibnamefont{Shen}}
  (\bibinfo{year}{2010}), \eprint{1008.3030}.

\bibitem[{\citenamefont{Gezerlis and Carlson}(2010)}]{Gezerlis:2009iw}
\bibinfo{author}{\bibfnamefont{A.}~\bibnamefont{Gezerlis}} \bibnamefont{and}
  \bibinfo{author}{\bibfnamefont{J.}~\bibnamefont{Carlson}},
  \bibinfo{journal}{Phys. Rev.} \textbf{\bibinfo{volume}{C81}},
  \bibinfo{pages}{025803} (\bibinfo{year}{2010}), \eprint{0911.3907}.

\end{thebibliography}

\end{document}